\documentclass[aps,amsfonts,pra,twocolumn,showpacs]{revtex4-1}
\usepackage{epsfig,amsmath,amssymb,bm,epsf,graphicx,psfrag,color,dsfont,natbib,subcaption,epstopdf}

\usepackage{bbold}
\usepackage{diagbox}

\usepackage{listings}

\lstnewenvironment{verbquote}[1][]
  {\lstset{columns=fullflexible,
           basicstyle=\ttfamily,
           xleftmargin=2em,
           xrightmargin=2em,
           breaklines,
           breakindent=0pt,
           #1}}
  {}

\def\bra#1{\langle#1\vert}
\def\ket#1{\vert#1\rangle}
\newcommand{\bc}{\begin{center}}
\newcommand{\ec}{\end{center}}
\newcommand{\be}{\begin{equation}}
\newcommand{\ee}{\end{equation}}

\newcommand{\outerproduct}[2]{|#1\rangle\langle #2|}

\newcommand{\shortexp}[1]{\mathrm{e}^{#1}}
\newcommand{\iden}{\mathds{1}}

\begin{document}

\title{Quantum algorithm for  spectral projection by measuring an ancilla iteratively}
\author{Yanzhu Chen}
\affiliation{C. N. Yang Institute for Theoretical Physics and Department of Physics and Astronomy, State University of New York at Stony Brook, Stony Brook, NY 11794-3840, USA}
\affiliation{Institute for Advanced Computational Science, State University of New York at Stony Brook, Stony Brook, NY 11794-5250, USA}
\author{Tzu-Chieh Wei}
\affiliation{C. N. Yang Institute for Theoretical Physics and Department of Physics and Astronomy, State University of New York at Stony Brook, Stony Brook, NY 11794-3840, USA}
\affiliation{Institute for Advanced Computational Science, State University of New York at Stony Brook, Stony Brook, NY 11794-5250, USA}
\date{\today}

\begin{abstract}
We propose a quantum algorithm for projecting  a quantum system to eigenstates of any Hermitian operator, provided one can access the associated control-unitary evolution for the ancilla and the system, as well as the measurement of the controlling ancillary qubit.  Such a Hadamard-test like primitive is iterated so as to achieve the spectral projection, and the distribution of the projected eigenstates obeys the Born rule. This algorithm can be used as a subroutine in the quantum annealing 
procedure by measurement to drive the system to the ground state of a final Hamiltonian, and we simulate this for quantum many-body spin chains.\end{abstract}

\maketitle

\section{Introduction}

The measurement postulate of quantum mechanics states that when measuring an observable 	$\hat{o}$, only its eigenvalues $o_n$  will be observed and the state of the system will be projected to the corresponding eigenstate $|o_n\rangle$, for which $\hat{o}|o_n\rangle=o_n|o_n\rangle$, immediately after the measurement. Furthermore, the Born rule prescribes the probability of such an outcome for  an initial quantum state $|\psi_0\rangle$  as $p_n=|\langle o_n|\psi_0\rangle|^2$. Whether one can derive the rule and hence remove it from the postulates of quantum mechanics is still of fundamental interest~\cite{Masanes2018}. 
From the perspective of quantum information processing, general construction of such spectral projection is also of practical importance.  For example, Ref.~\cite{Poulin2018} constructs a quantum walk approach to achieve this and emphasizes its utility in carrying out a key step of the quantum simulated annealing (QSA) algorithm for optimization problems~\cite{Somma2008}. The latter can be used as an alternative to the adiabatic quantum computation (AQC)~\cite{Adia1,Adia2}. In fact, the standard quantum phase estimation (QPE)~\cite{NielsenChuang} and its variants~\cite{Kitaev2002,Aspuru-Guzik2005,Dobsicek2007} can also achieve approximate spectral projection when the system is not in an eigenstate.

	The QPE is crucial in many quantum information processing  applications~\cite{NielsenChuang}, including factoring and, more relevant to the present paper, the quantum-walk spectral measurement in Ref.~\cite{Poulin2018}, as well as  related methods for preparing a thermal Gibbs state~\cite{Poulin2009,Temme2012,Yung2012,Moussa2019}. The standard QPE uses  ${\cal O}(t_g)$ controlled unitary gates of the form $c-U^{2^k}$ (for $k=0$ to $t_g-1$) to encode the $t_g$ binary digits of  the phase value (in unit of $2\pi$) and it requires ${\cal O}(t_g^2)$ gates in the inverse quantum Fourier transform to retrieve the phase~\cite{NielsenChuang}. Regarding the accuracy of QPE, in order to have  the phase  accurate in $m$ binary digits with the success probability of  at least $1-\epsilon$, the total number of ancillary qubits needed is $t_g=m+\log(2\epsilon+1/2\epsilon)$~\cite{NielsenChuang}. In other words, using $t_g$ ancillary qubits allows the phase value to be accurate in $t_g-\log(2\epsilon+1/2\epsilon)$ binary digits. The accuracy in the phase is thus limited by the number of available ancillas employed in representing the value of the phase,  and when used as spectral projection subroutine, the eigenstate the system is projected by the QPE to is only approximate. The unitary $U$ may be implemented by $e^{-i \hat{o} \Delta t}
$, and in the QPE, the power in the unitary $U$ needs to go as large as  $2^{t_g-1}$; equivalently, the timing $\Delta t$ needs to be made accurate to $2^k$ (for $k=0$ to $t_g-1$).  Maintaining the stability of $U$ and coherence of the quantum register  when carrying out the QPE is important for noisy intermediate-scale quantum processors. 
	
	Here, we apply a simple iterative approach to achieve the spectral projection of an associated observable $\hat{o}$, and in each step of the iteration only one ancilla is used as the control to enact a unitary evolution (${\rm c-}e^{-i \Delta t \hat{o}}$) on the system, conditioned on the ancillary state being $|1\rangle$. Then only the ancilla is measured in the Pauli X basis. After sufficient number of steps have been carried out (see below), the system is projected to an eigenstate of the operator $\hat{o}$. We  demonstrate by numerical simulations that our procedure can lead to spectral projection by varying the parameter $\Delta t$ and the ancilla's state parameter.
	
	To understand that repeated application of the primitive eventually leads to spectral projection, we provide two perspectives. First, we  show that on average the energy variance of the system will decrease; see Eq.~(\ref{eqn:dVarNew}). If the energy variance decreases to zero, then an eigenstate is reached.
	Second, an intuitive picture of our procedure emerges:  at each step, the measurement of the ancillary qubit gives rise to a random walk in the operator action, i.e.  with either $e^{\hat{Q}_0}$ or $e^{\hat{Q}_1}$ acting on the system. The choice of which operators depends on the measurement outcome; see Fig.~\ref{fig:oneIteration} below. The key notable difference from the conventional random walk is that the outcome probability is state dependent. However, we calculate the average random-walk action $p_0\hat{Q}_0+p_1\hat{Q}_1$ per step that is valid in the small $\Delta t$ limit, and find that it leads to a map, see Eq.~(\ref{eqn:decayNew}), that when repeated will drive the system to an eigenstate.  Both viewpoints validate that our procedure can lead to spectral projection, as eigenstates have no energy  (or observable-value) variance and are fixed points of the iterative procedure.
	
	We emphasize that the time $\Delta t$ here, unlike in the QPE, does not need to be exactly of the form $2^k$. Thus, in some sense the protocol for spectral projection does not require exact timing and can tolerate fluctuations and imprecision in timing. In addition, the range of $\Delta t$ used needs not span over many orders of magnitudes related to the accuracy of the QPE, i.e., $\max\{\Delta t\}/\min\{\Delta t\}$ can be much smaller than $2^{t_g-1}$. Moreover, the ancilla state does not need to be in the $|\pm\rangle$ state right before the controlled unitary and it can be in almost any pure state. As seen below, we can also used a fixed  $\Delta t$ in our procedure  to achieve the spectral projection.

	Given that spectral projection can be achieved, one immediate question is what governs the distribution of the projected eigenstates. For this we show that the distribution of this eigenstate projection obeys the Born rule. Fundamentally, our algorithm can be regarded as a procedure to achieve the effect described in the measurement postulate. 
	As an application, we simulate the use of our spectral projection algorithm in two spin-chain models, and demonstrate that ground states at different transverse field strengths can be successfully obtained, when there is a gap in the Hamiltonian throughout the parameter range of interest.  
	
	Our initial motivation for this study comes from the incentive to devise a simple quantum version of Lanczos algorithm. An approach was recently proposed in Ref.~\cite{Motta2019} by implementing an effective unitary evolution $e^{-i h_{\rm eff} \Delta \tau}$  to simulate the effect of imaginary time evolution $e^{- h\Delta\tau}$ on a quantum state. We wish to develop an alternative approach that does not require the searching of the effective Hamiltonian $h_{\rm eff}$. However, we could not make the procedure to work due to high-order effect, and we describe such a failed attempt in the Appendix. However,  it was by analyzing this that  leads us to the spectral projection algorithm and the understanding why the attempt failed.
	
	The remainder of the paper is organized as follows. In Sec.~\ref{sec:prelim} we discuss a primitive that slightly generalizes the Hadamard test by using a general ancillary state. By repeating this primitive with sufficient number of times, we argue that it will project the system to an eigenstate.  In  Sec.~\ref{sec:simulations} we describe the approach to classically simulate the above procedure and verify by simulations that it indeed leads to an eigenstate or spectral projection algorithm. There, we use random Hermitian matrices for illustration and also demonstrate that such spectral projection obeys the Born rule for the final distribution of projected eigenstates.  In Sec.~\ref{sec:physical} we give illustrations of our spectral projection algorithm using the quantum transverse-field Ising spin chain.   In Sec.~\ref{sec:decoherence} we discuss the effect of decoherence. In Sec.~\ref{sec:subroutine} we illustrate the use of our spectral projection algorithm in the quantum annealing for two different spin chains.
	Finally, in Sec.~\ref{sec:conclusion}, we make some concluding remarks.
	
	\begin{figure}[t]
	  (a)\\ 	\includegraphics[width=0.45\textwidth]{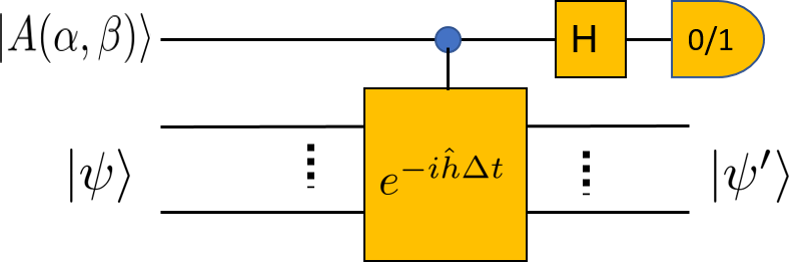}\\
\vspace{0.1cm}	 (b)\\ \includegraphics[width=0.45\textwidth]{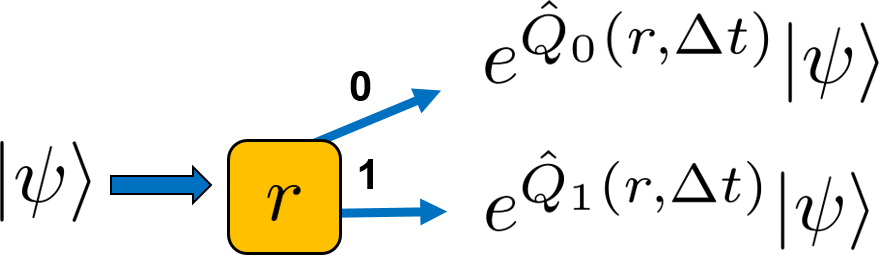}
		\caption{Basic picture of  our algorithm. (a) The primitive: one ancilla  is used as the control qubit for the control unitary, which is jointly applied to the ancilla and the system, ${\rm cU}=|0\rangle\langle 0|\otimes I + |1\rangle\langle 1|\otimes e^{-i \Delta t \,\hat{h}}$, followed by a measurement on the ancilla in the $X\equiv\sigma^x$ basis. (b) Summary of the action on the input system state: $|\psi'_m\rangle\sim e^{\hat{Q}_m}|\psi\rangle$. This leads to a random walk picture for the algorithm.}
		\label{fig:oneIteration}
	\end{figure}
\section{The primitive and the alogrithm for spectral projection }
\label{sec:prelim}

The primitive that our algorithm is based on is similar to the Hadamard test and will be described below. The algorithm itself is a repeated application of such a primitive. We will provide analysis to support that our algorithm can achieve spectral projection.
\subsection{The Hadamard test and the primitive} 	
The basic idea of our approach is to  entangle a system with an ancilla qubit prepared in a certain state, and then measure the ancilla in a chosen basis, similar to the so-called Hadamard test. This is commonly used in many quantum information processing protocols~\cite{NielsenChuang}. We will describe a slightly varied primitive, in which the ancilla needs not be in the $|+\rangle$. 

Let the system be 
in an initial state $\ket\psi$ and an ancilla in  $| A\rangle = \alpha\ket0+\beta\ket1$ with $|\alpha|^2+|\beta|^2=1$. We entangle the ancilla (as the control) and the system (as the target) by the controlled operation ${c-U}\equiv\outerproduct{0}{0}\otimes\iden+\outerproduct{1}{1}\otimes U$, where $U=\exp{(-i\hat{h}\Delta t)}$ is the unitary evolution under a Hamiltonian $\hat{h}$ within a duration $\Delta t$. We then measure the controlling ancillary qubit in the basis $(\ket0 \pm \shortexp{i\xi}\ket1)/\sqrt{2}$, with the $\pm$ associated with the measurement outcome $m=0$ or $1$, respectively.  This is equivalent to measuring the observable $\cos(\xi)\sigma_x +\sin(\xi)\sigma_y$ on the ancilla.
We shall see below that we can take $\xi=0$ without loss of generality, and thus the measurement will correspond to the Pauli X basis, and the primitive is illustrated in Fig.~\ref{fig:oneIteration}.

The measurement of the ancilla then collapses the system to the unnormalized state:
	\begin{align}
|\psi'_m\rangle&=\frac{1}{\sqrt{2}}
	[\bra0+(-1)^m\shortexp{-i\xi}\bra1] \left[\alpha\ket0\ket\psi+\beta\ket1 U\ket\psi\right] \nonumber \\
	&=\frac{1}{\sqrt{2}}[\alpha+(-1)^m\shortexp{-i\xi}\beta e^{-i\hat{t}\Delta t}] \ket\psi, \label{eqn:post}
	\end{align}
	and the corresponding probability of obtaining the outcome $m$ is
	\begin{align}
	\label{eq:p_m}
	p_m=||\psi'_m||^2=\frac{1}{2}\big[1+2(-1)^m {\rm Re}(\alpha^*\beta e^{-i\xi} \langle \psi|e^{-i\hat{h}\Delta t}|\psi\rangle)\big]. 
	\end{align}
Here we see that the phase factor $e^{-i\xi}$ from the measurement basis can be absorbed into the ancilla's initial state parameter $\beta$, and thus we can set $\xi=0$ from now on without loss of generality, resulting in the ancilla measurement to the fixed the Pauli X, whose eigenstates are  simply $|\pm\rangle\equiv (|0\rangle\pm|1\rangle)/\sqrt{2}$.

\medskip 
\noindent {\bf Eigenstates are fixed points of the primitive}.
It is easy to see that  for any eigenstate $|E_j\rangle$ with eigenenergy $E_j$, the post-measurement state is still $|E_j\rangle$, but the probability of getting the $m$-th outcome is
\begin{eqnarray}
p_{j,m}&=&\frac{1}{2}|\alpha+(-1)^m \beta e^{-i E_j\Delta t}|^2\\
&=&\frac{1}{2}[1+2(-1)^m {\rm Re}(\alpha^*\beta e^{-iE_j\Delta t})].
\end{eqnarray}
The probabilities for  `0' and `1' outcomes add up to unity: $p_{j,0}+p_{j,1}=1$. 
Moreover, their difference $p_{j,0}-p_{j,1}= 2 {\rm Re}(\alpha^*\beta e^{-iE_j\Delta t})=2|\alpha\beta|\cos(\phi-E_j\Delta t)$ can be used to determine $E_j\Delta t$ up to an overall sign and multiples of $2\pi$, where $\alpha^*\beta=|\alpha\beta|e^{i\phi}$. To uniquely determine $E_j$, one can use a different set of  $(\alpha,\beta)$ and $\Delta t$ to obtain different distributions for estimation. Note that in order to achieve optimal determination we can maximize $|\alpha\beta|$, which is achieved when $|\alpha|=|\beta|=1/\sqrt{2}$ and corresponds to using an ancillary state $|A\rangle=(|0\rangle + e^{i\phi}|1\rangle)/\sqrt{2}$. The choice of $|+\rangle$ is the typical ancillary state in the Hadamard test.

Suppose we have two different energy eigenstates with distinct energies $E_j\ne E_k$,  generically the two distributions are different, $p_{j,m}\ne p_{k,m}$, unless the choice of $(\alpha,\beta)$ and $\Delta t$ coincidentally make  ${\rm Re}(\alpha^*\beta e^{-iE_j\Delta t})= {\rm Re}(\alpha^*\beta e^{-iE_k\Delta t})$. Hence, by accumulating enough statistics, one can determine whether the two eigenstates have the same energy or not.   One can use different `distance' measures, such as the relative entropy to quantify the distinguishability. Quantities such as the Chernoff bound can also be used to quantify the likelihood of deviating from the average values and thus the degree of distinguishability for a finite number of measurements performed.

For the system's initial state being $|\psi\rangle=\sum_j c_j^{(0)} |E_j\rangle$, then the probability of getting outcome $m$ in the ancilla's measurement can be shown to be
$p_m^{(0)}= \sum_j |c_j^{(0)}|^2 p_{j,m}$, which is a convex mixture of the extremal distributions $p_{j,m}$ from the eigenstates. This means that  the knowledge of the distribution ${p}_m^{(0)}$ is not sufficient to infer uniquely the compositions of the eigenstates. If we are given only a copy of $|\psi\rangle$ and if it is not in an eigenstate, then it is not possible to estimate $p_m^{(0)}$ by any measurement.

\medskip \noindent {\bf Energy change}.

First, we can ask how much the energy has changed after one such a primitive step:
$\Delta E_{(m)}\equiv  \langle \tilde{\psi'}_m|\hat{h}|\tilde{\psi'}_m\rangle -  \langle \psi|\hat{h}|\psi\rangle$, where $|\tilde{\psi'}_m\rangle\equiv |\psi'_m\rangle/\sqrt{p_m}$ is the normalized post-measurement state.  By using the expressions for the post-measurement state $|\psi'_m\rangle$~(\ref{eqn:post}) with $\xi=0$ and the probability $p_m$~(\ref{eq:p_m}), we can calculate $\Delta E_{(m)}$ explicitly and arrive at
 (see Appendix~\ref{app:derivations} for derivations)
\begin{equation}
\Delta E_{(m)}=\frac{2(-1)^m\big({\cal R}_h-\langle h\rangle {\cal R}_1\big)}{1+2(-1)^m {\cal R}_1}
\end{equation}
where parameters ${\cal R}_1$ and ${\cal R}_h$ are defined as
\begin{eqnarray}
\label{eq:R1}
{\cal R}_1&\equiv& {\rm Re}\big(\alpha^*\beta \langle \psi|e^{-i\hat{h}\Delta t}|\psi\rangle\big), \\
{\cal R}_h&\equiv& {\rm Re}\big(\alpha^*\beta \langle \psi|e^{-i\hat{h} \Delta t}\hat{h}|\psi\rangle\big). \label{eq:Rh}
\end{eqnarray}
In order to obtain some intuition of the above expression, we can expand it to the first nonvanishing order. We find that for ${\rm Im}(\alpha^*\beta)\ne 0$, the lowest nonvanishing contribution occurs  at
 the first order in $\Delta t$,
\begin{equation}
\label{eq:dEgeneral0}
\Delta E_{(m)}=\frac{2(-1)^m {\rm Im}(\alpha^*\beta)}{1+2(-1)^m {\rm Re}(\alpha^*\beta)}\langle\Delta h^2\rangle \Delta t,
\end{equation}
where the expectation $\langle \cdots \rangle$ is evaluated w.r.t. $|\psi\rangle$, e.g. $\langle \hat{h}\rangle\equiv \langle\psi|\hat{h}|\psi\rangle$, and the $\langle(\Delta h)^2\rangle\equiv\langle\psi|\hat{h}^2|\psi\rangle-\langle\psi|\hat{h}|\psi\rangle^2$ is the energy variance of the state $|\psi\rangle$. 
Thus, generically the change in the energy after one step is proportional to the energy variance before the application of the primitive.

We note that, however, when ${\rm Im}(\alpha^*\beta)= 0$, the change in energy is in the second order, 
\begin{equation}
\Delta E_{(m)}=-\frac{{\rm Re}(\alpha^*\beta)\big(\langle\hat{h}^3\rangle -\langle\hat{h}^2\rangle \langle\hat{h}\rangle\big)(\Delta t)^2}{1+2(-1)^m {\rm Re}(\alpha^*\beta)}.
\end{equation}
Of course, the exception is when ${\rm Re}(\alpha^*\beta)=\pm 1/2$, which corresponds to the case of the ancillary state being $|\pm\rangle$.  
In the case of using $|+\rangle$ of the ancilla, the change in energy to the first nonvanishing contribution is
\begin{equation}
\label{eq:dE2+}
\Delta E=\left\{ \begin{array}{ll}
\frac{1}{4}(\langle\hat{h}^3\rangle -\langle\hat{h}^2\rangle \langle\hat{h}\rangle)(\Delta t)^2, & m=0,\\
\phantom{} & \phantom{}\\
\frac{\langle\hat{h}^3\rangle}{\langle\hat{h}^2\rangle} - \langle\hat{h}\rangle, & m=1,
\end{array}\right.
\end{equation}
The above two outcomes are switched, if the ancilla's state is $|-\rangle$.

It is interesting to observe that in general  the energy will always change if $|\psi\rangle$ is not an eigenstate, except when the ancilla's state satisifies  ${\rm Im}(\alpha^*\beta)= 0$ (e.g. in the $|\pm\rangle$ state)  and the system satisfies $\langle\hat{h}^3\rangle -\langle\hat{h}^2\rangle \langle\hat{h}\rangle=0$, then the energy will not change.

\subsection{The algorithm}
The algorithm is simply a procedure that repeats the above primitive many times. 
In each step, the ancilla parameters $(\alpha,\beta)$ and the duration $\Delta t$ can be different. 
In the following, we provide argument to support that our algorithm indeed will achieve spectral projection, by analyzing two quantities that characterize the average effect, in terms of the energy variance and the average action of a random walk.

Before we proceed to the analysis, we need to first ask the question: how do we know the algorithm has produced a converged eigenstate? Assume that the system converges to an eigenstate $|E_j\rangle$. Applying the $c-U$ gate to the ancilla  (initially in $\alpha|0\rangle+\beta|1\rangle$) and the system leaves the system intact but changes the relative phase in the ancilla: $\alpha|0\rangle +\beta e^{-i E_j\Delta t}|1\rangle$. 
Following the idea in the so-called eigenstate witness method~\cite{Santagati2018},  one can perform quantum state tomography on the identically prepared ancillary qubit after applying the control unitary. A single-qubit tomography involves measurement in Pauli X, Y, and Z bases, as for a general one-qubit mixed state $\rho_{a}=(I+\sum_{i=x,y,z} r_i \sigma_i)/2$, its parameters can be obtained from measurement: $r_i={\rm Tr}(\rho_a \sigma_i)$. In our algorithm, we perform X measurement, but measurement in Y can also achieve spectral projection,  as we have argued that the measurement phase $\xi$ in $|0\rangle \pm e^{i\xi} |1\rangle$ was conveniently aborbed in the ancilla parameter $\beta$. We can also perform Pauli $Z$ measurement, which will project the ancilla and the system to $|0\rangle|\psi\rangle$ or $|1\rangle e^{-i\hat{h}\Delta t}|\psi\rangle$, and it does not affect the spectral projection. If the state tomography shows that it remains in a pure state, then the system must be in an eigenstate and hence it is converged. One can use the purity of the resultant ancillary state as a measure for convergence. Moreover, from the tomography, the quantity $E_j\Delta t$ can also be determined (up to multiple of $2\pi$). By using different sets of $\Delta t$, one can then uniquely determine $E_j$.

\medskip 
\noindent {\bf Perspectives from probability distribution}.
As we shall see below, successive coupling of  individual ancillas with the system and measurement on ancillas help to drive the system to an eigenstate, therefore arriving at the final distribution  $p_m=p_{j^*,m}$ for some $j^*$ labeling the eigenstate $|E_{j^*}\rangle$. The procedure gives rise to a sequence of 0/1 outcomes, namely a bit string $0,1,1,0,\dots$ from the ancillas' measurement.   From the persective of probability distribution, it starts with $p_m^{(0)}$, and after successive application of the primitive, the distribution flows: $p_m^{(0)}\rightarrow p_m^{(1)}\rightarrow \cdots p_m^{(n)}$,  and after $n$ steps, $p_m^{(n)}$ will be close to some $p_{j^*,m}$. (In terms of coins, there are $n+1$ different coins.) We stress, however, that in the process we cannot obtain the distributions $p_m^{(k)}$ from measurement as we are given only one copy of the system, but we only obtain a sequence of 0 and 1. From this sequence we can only estimate the average probability of getting 0 and 1, i.e. $\overline{p_m}$.  In the case of the eigenstates, such average distribution can be used to distinguish whether two given eigenstates have different energy or not (as the eigenstate does not change, and hence the `coins' are identical, as we have discussed previously). However, for an arbitrary initial state of the system, does the knowledge of $\overline{p_m}$ guarantee the projection? 

Given that our algorithmic procedure enables the projection to eigenstates (as argued and numerically demonstrated below), then from the bit string and the knowledge of the initial state of the system, one can indeed infer  the distributions $p^{(k)}_m$, as well as whether and what energy eigenstate is arrived and what the corresponding eigenenergy is by direct classical simulations. However, for large system sizes, classical simulations will not be possible. How do we argue that our procedure indeed leads to spectral projection? How do we explain that in the limit of long seqence $\overline{p_m}$ will eventually flow to an extremal or fixed-point distribution $p_{j^*,m}$?
In the following, we provide two physically motivated approaches to understand the spectral projection. 

\smallskip\noindent {\bf Energy variance}. If the energy variance of a quantum state is zero, the state is an energy eigenstate, i.e., $V_E(\psi)=0 \leftrightarrow \hat{h}|\psi\rangle=E |\psi\rangle$, where $V_E(\psi)\equiv\langle\psi| (\Delta h)^2|\psi\rangle=\langle\psi |\hat{h}^2|\psi\rangle-
\langle\psi |\hat{h}|\psi\rangle^2$. Thus, energy variance is an important indicator to how close the state has converged to an eigenstate.
Carrying out the primitive yields the outcome `0' with probability $p_0$ and the normalized post-measurement system state $|\tilde{\psi}'_0\rangle\equiv |\psi'_0\rangle/\sqrt{p_0}$, and the outcome `1' with probability $p_1$ and the normalized post-measurement system state $|\tilde{\psi}'_1\rangle/\equiv |\psi'_1\rangle/\sqrt{p_1}$. Given the probabilistic nature due to measurement, it is thus natural consider the average change of the energy variance after one step:
\begin{eqnarray*}
\overline{\delta V_E}&\equiv& \big[p_0 \big(\langle\tilde{\psi}'_0| \hat{h}^2|\tilde{\psi}'_0\rangle - \langle\tilde{\psi}'_0| \hat{h}|\tilde{\psi}'_0\rangle^2\big)\big. \\
&&+\big. p_1\big(\langle\tilde{\psi}'_1|\hat{h}^2|\tilde{\psi}'_1\rangle-\langle\tilde{\psi}'_1|\hat{h}|\tilde{\psi}'_1\rangle^2\big)\big] - \langle\psi| (\Delta h)^2|\psi\rangle.\nonumber
\end{eqnarray*}
Since the expectation value of $\hat{h}$ and its function such as $\hat{h}^2$ are conserved, the above change can be simplified to be
\begin{equation}
\overline{\delta V_E}= \langle\psi|\hat{h}|\psi\rangle^2-\big[p_0 \langle\tilde{\psi}'_0| \hat{h}|\tilde{\psi}'_0\rangle^2+ p_1\langle\tilde{\psi}'_1|\hat{h}|\tilde{\psi}'_1\rangle^2\big].
\end{equation}

By using the expressions for the post-measurement state~(\ref{eqn:post}) and the probability~(\ref{eq:p_m}), we can calculate $\overline{\delta V_E}$ explicitly and arrive at (see Appendix~\ref{app:derivations} for derivations)
\begin{equation}
\overline{\delta V_E}=\frac{-4}{1-4 {\cal R}_1^2}\big({\cal R}_1\langle h\rangle - {\cal R}_h\big)^2,
\end{equation}
where the two parameters ${\cal R}_1$ and ${\cal R}_h$ are defined previously as in Eqs.~(\ref{eq:R1}) and~(\ref{eq:Rh}), respectively.
Given that $|\alpha^*\beta|\le 1/2$, the parameter $R_1$ satisfies $R_1^2\le 1/4$. Thus, $\overline{\delta V_E}\le 0$ and generically $\overline{\delta V_E}<0$.

In order to obtain some intuition about $\overline{\delta V_E}$, we can expand it in series of $\Delta t$, and we find that when ${\rm Im}(\alpha\beta)\ne 0$, it is nonvanishing at the second order:
\begin{eqnarray}
\label{eqn:dVarNew}
\overline{{\delta V_E}}=-\frac{4\,{\rm Im}(\alpha^*\beta)^2 }{1- 4{\rm Re}(\alpha^*\beta)^2}\big(\langle\psi|(\Delta h)^2|\psi\rangle\big)^2(\Delta t)^2.
\end{eqnarray}
The  factor, $c(\alpha,\beta)\equiv {{\rm Im}(\alpha^*\beta)^2}/\big[{1-4 {\rm Re}(\alpha^*\beta)^2}\big] $ is maximized with a value $1/4$ when $\beta/\alpha=e^{i\phi}$ and $|\alpha|=|\beta|=1/\sqrt{2}$.  
Such a choice represents a maximum `rate' of change in the energy variance. 
Moreover, the change is also proportional to the square of the energy variance of the system's state before carrying out the primitive.

We note that, however, when ${\rm Im}(\alpha^*\beta)= 0$,
\begin{equation}
\overline{{\delta V_E}}=-\frac{{\rm Re}(\alpha^*\beta)^2\big(\langle\hat{h}^3\rangle -\langle\hat{h}^2\rangle \langle\hat{h}\rangle\big)^2(\Delta t)^4}{1-4 {\rm Re}(\alpha^*\beta)^2},
\end{equation}
except when $\alpha=1/\sqrt{2}$ and $\beta=\pm1/\sqrt{2}$, the average change in the energy variance is
\begin{eqnarray}
\label{eqn:dVarNew+-}
\overline{{\delta V_E}}=-\frac{\big(\langle \hat{h}^3\rangle -\langle\hat{h}^2\rangle \langle\hat{h}\rangle\big)^2}{4\langle \hat{h}^2\rangle}(\Delta t)^2.
\end{eqnarray}
We observe that the quantity $\langle \hat{h}^3\rangle -\langle\hat{h}^2\rangle \langle\hat{h}\rangle$ has previously appeared in the change of the energy~(\ref{eq:dE2+}).

The above analysis suggests that we should usually choose ancillary parameters such that  ${\rm Im}(\alpha^*\beta)\ne 0$, except when $\alpha=1/\sqrt{2}$ and $\beta=\pm1/\sqrt{2}$, so as to make the average energy variance decrease in ${\cal O}(\Delta t^2)$. If the energy variance continues to decrease closely to zero, then an energy eigenstate is approached.  We remark that as demonstrated below it is not necessary to use the same ancillary state and time duration $\Delta t$ in every step of the procedure. Varying ancilla's state away from $|\pm\rangle$ can be useful to avoid the system state to get stuck in  states that have $\langle \hat{h}^3\rangle -\langle\hat{h}^2\rangle \langle\hat{h}\rangle=0$. See also below in Sec.~\ref{sec:physical} for further discussions on this.

\smallskip\noindent
{\bf Random-walk approach}. As the procedure outputs a pure state if the input is also pure, a question arises as to how we can analytically understand how  the system is eventually driven to an eigenstate?
Let us analyze the post-measurement states $|\psi'_m\rangle$ by expanding it to the second order in $\Delta t$,
	\begin{eqnarray}
|\psi'_m\rangle\approx \frac{\alpha+(-1)^m\beta}{\sqrt{2}}\left[1+\frac{-i\hat{h}\Delta t -\frac{1}{2}(\hat{h}\Delta t )^2}{1+(-1)^m\alpha/\beta}\right]|\psi\rangle.\nonumber
	\end{eqnarray}
	We can rewrite the above equation to find the exponentiated action on $|\psi\rangle$, i.e.,
	$|\psi'_m\rangle\sim e^{\hat{P}_m} |\psi\rangle$ and ignore the overall constant. As shown in Appendix~\ref{app:derivations},
we find that to the second order in $\Delta t$
	\begin{eqnarray}
	\hat{P}_m= \frac{-i\hat{h}\Delta t -\frac{1}{2}(\hat{h}\Delta t )^2}{1+(-1)^m\alpha/\beta}+\frac{1}{2}\frac{(\hat{h}\Delta t )^2}{[1+(-1)^m\alpha/\beta]^2}.
	\end{eqnarray}
	As $\hat{P}_m$ is a polynomial of $\hat{h}$,  one can separate it into two commuting parts: one Hermitian and the other anti-Hermitan, $\hat{P}_m= (\hat{P}_m + \hat{P}_m^\dagger)/2 + (\hat{P}_m-\hat{P}_m^\dagger)/2=: \hat{Q}_m + i\hat{ R}_m$. As the part $i\hat{R}_m$ is anti-Hermitian, its corresponding action $e^{i \hat{R}_m}$ is a unitary, and it does not modify the relative weight in the decomposition of energy eigenstates, so we can ignore it when we consider  eigenstate projection. Thus, we focus on 	$|\psi'_m\rangle\sim e^{\hat{Q}_m}|\psi\rangle$, where $\hat{Q}_m= (\hat{P}_m + \hat{P}_m^\dagger)/2$.

After a long sequence of iterations, we will have a long product of operators $e^{\hat{Q}}$'s (which commute with one another)  acting on the initial state $|\psi\rangle$, such as
\begin{equation}
 e^{\hat{Q}_1(\alpha,\beta,\Delta t)}e^{ \hat{Q}_0(\alpha,\beta,\Delta t)}e^{\hat{Q}_1(\alpha,\beta,\Delta t)}\cdots e^{ \hat{Q}_0(\alpha,\beta,\Delta t)},
\end{equation}
which looks like a  sequence of `random walk' using the two operators in the exponent. 
However, the key difference from a  typical random walk is that there is a quantum state that changes after every step and the probability of moving to the left or right $p_{0/1}$ is state dependent, as in Eq.~(\ref{eq:p_m}).

Here, as an approximation for the average action $e^{\overline{\hat{Q}(\alpha,\beta,\Delta t)}}$, which is valid in the limit $\Delta t\rightarrow 0$, we ignore the subsequent state dependence and use the initial $p_{0/1}(\psi)$ to calculate the average in the exponent:
$p_0(\psi)\cdot \hat{Q}_0 + p_1(\psi) \cdot \hat{Q}_1$, and we arrive at 
\begin{align}\sum_{m=0,1}p_m \hat{Q}_m=-\frac{{\rm Im}(\alpha^*\beta)^2 \Delta t^2}{1-4 {\rm Re}(\alpha^*\beta)^2}[(\hat{h}-\langle \hat{h}\rangle)^2 - \langle \hat{h}\rangle^2].
\end{align}
Thus, the average one-step action gives rise to a map on the system:
\begin{equation}
\label{eqn:decayNew}
|\psi\rangle\rightarrow e^{-c(\alpha,\beta) \Delta t^2 (\hat{h}-\langle \hat{h}\rangle)^2 }|\psi\rangle.
\end{equation}
 The  factor $c(\alpha,\beta)\ge 0$ is defined earlier and is maximized with a value $1/4$ when $\beta/\alpha=e^{i\phi}$ and $|\alpha|=|\beta|=1/\sqrt{2}$.  This represents the optimal choice of ancillary parameters to maximize the converge rate, consistent with results presented earlier. 

The meaning of the above equation is that the procedure tends to suppress components of eigenstates that have eigenvalues further away from $h_\psi\equiv\langle \psi|\hat{h}|\psi\rangle$.  As one repeatedly applies the primitive, the state $|\psi\rangle$ itself will change and hence so will the expectation value $\langle\psi|\hat{h}|\psi\rangle$, with  the latter eventually approaching the energy eigenvalue and the system state approaching the corresponding eigenstate. 
The random-walk analysis gives similar conclusion as that by the change in the average energy variance. We note that when ${\rm Im}(\alpha^*\beta)=0$ and the ancilla's state not being $|\pm\rangle$, we need to carry out the expansion to the fourth order, but we do not perform the calculation here.

\section{Procedure for classical simulations }
\label{sec:simulations}
 The primitive looks similar to the Hadamard test and consists a controlled-unitary action on the ancilla and the system, as well as a subsequent measurement on the ancillary qubit. Since the effect is to update the state vector of the system, for classical simulations of this process, we only need to compute two (un-normalized) wave functions $|\psi_m^{(k)}\rangle$ and their norm squares $p_m^{(k)}\equiv\langle \psi_m^{(k)}|\psi_m^{(k)}\rangle$ at each step, say, $k$-th, 
\be
|\psi_m^{(k)}\rangle=\frac{1}{\sqrt{2}}\left[\alpha_k |\psi^{(k-1)}\rangle+ (-1)^{m}\beta_k U_k|\psi^{(k-1)}\rangle\right],
\ee
given the  state, $|\psi^{(k-1)}\rangle$,  of the system from the end of the previous step, the parameters $\alpha_k$ and $\beta_k$, and the unitary $U_k(\Delta t_k)=e^{-i \Delta t_k \hat{h}}$.

One then decides  to update the state $|\psi^{(k)}\rangle=|\psi^{(k)}_m\rangle/\sqrt{p_m}$ by choosing $m=0$ or $m=1$ with  probability  $p_m^{(k)}$.
With a suitable choice of $\{(\alpha_k,\beta_k)\}$ and $\{\Delta t_k\}$, the long-iterated state $|\psi^{(k\gg 1)}\rangle$ will converge to some eigenstate $|E_n\rangle$, as illustrated below.

Simulating this procedure for spectral projection also provides us  a quantum-inspired classical algorithm to obtain (randomly) excited states, whose accuracy does not depend on other lower lying levels. The costly part is applying $e^{-i \hat{h} \Delta t}$ to a state vector. However, for the purpose of a short-range Hamiltonian, one can use the Trotter decomposition and the individual $e^{-i\hat{h}_j\Delta t}$ from $\hat{h}=\sum_j \hat{h}_j$. Tensor-network representations can also be useful.

\begin{figure*}
	(a)	\includegraphics[width=0.45\textwidth]{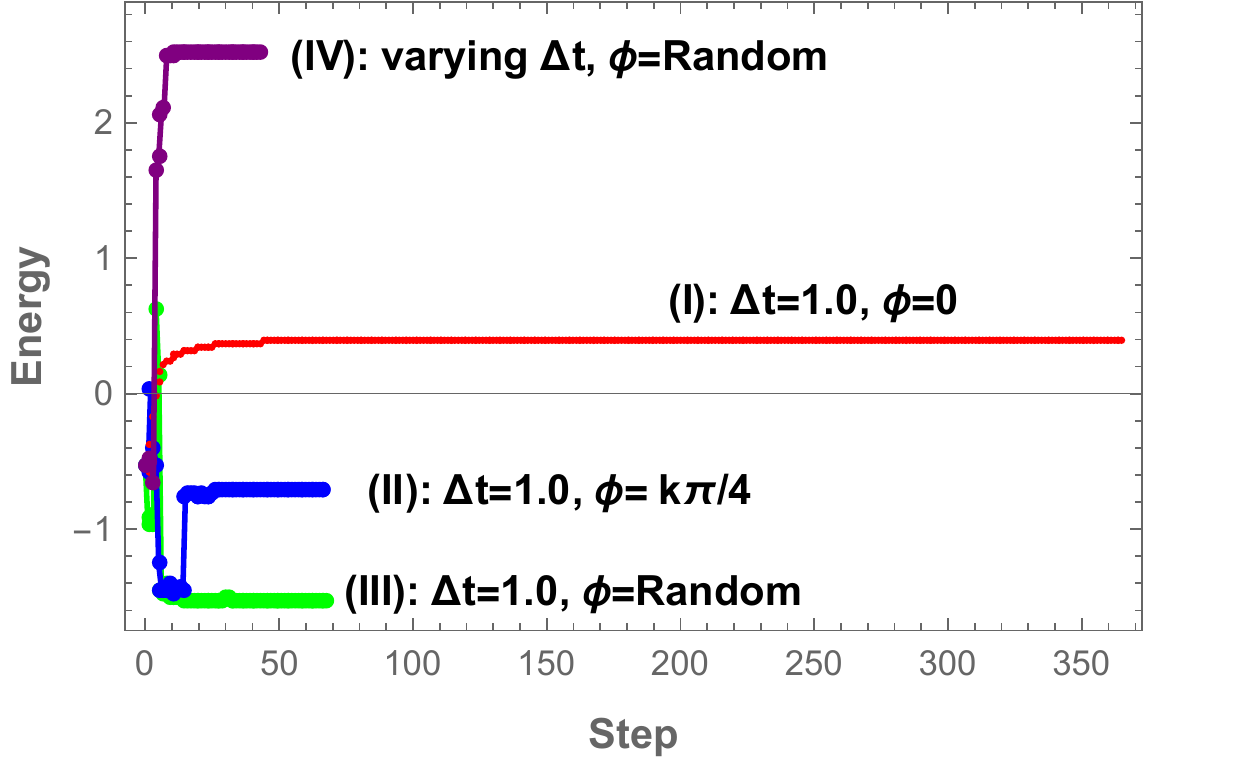}
	(b)	\includegraphics[width=0.47\textwidth]{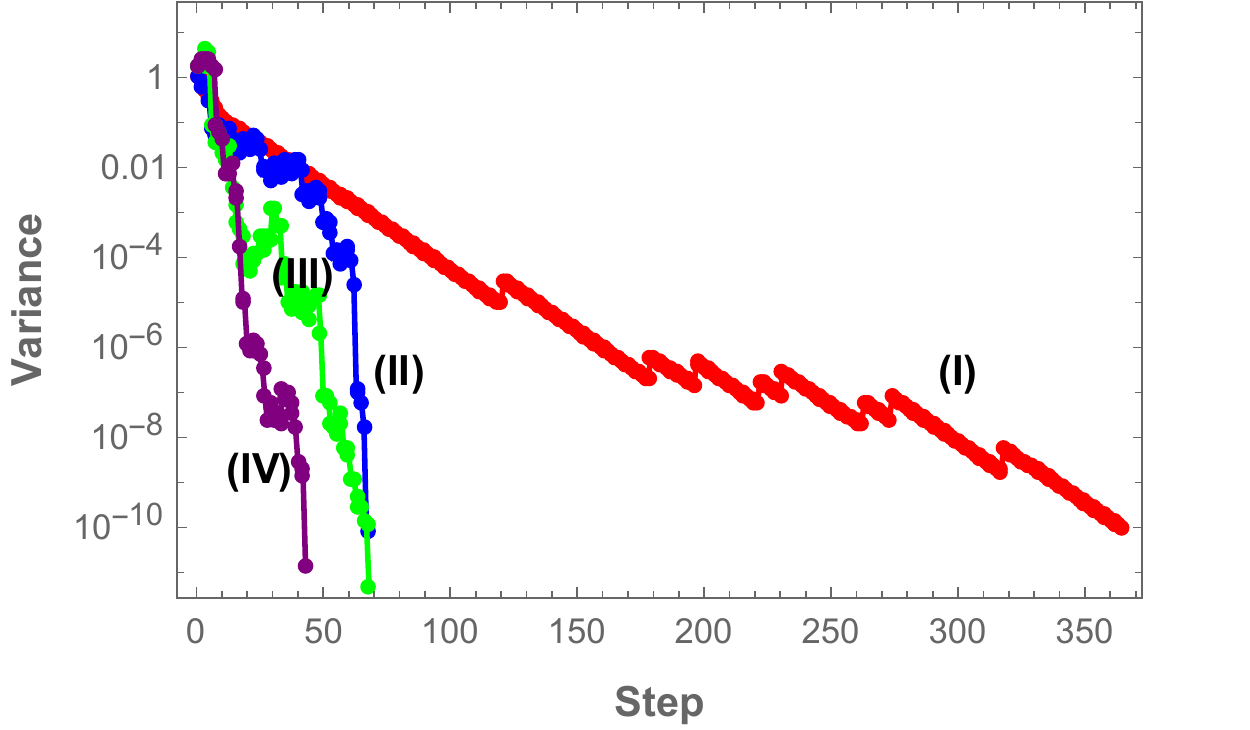}
		\caption{The iteration procedure using various choices of $\Delta t$ (arbitrary unit) and $\phi$. See the main text for detailed discussions of the four types of choices (I) to (IV). In (a) the energy is in an arbitrary unit and the values are recorded in the iteration; in (b) the energy variances (also in an arbitary unit) are recorded.  
		 }
		\label{fig:QVNew}
	\end{figure*}
To obtain the entire set of eigenstates, we need to simulate the spectral projection as many times as the Hilbert space dimension.  One can start with the system in an arbitrary initial $|\psi_0\rangle$ state. Run the procedure to obtain some eigenstate $|\phi_1\rangle$, then subtract the portion of $|\phi_1\rangle$ from $|\psi_0\rangle$: $|\psi^{[1]}\rangle= |\psi_0\rangle- \langle \phi_1 |\psi_0\rangle |\phi_1\rangle$ and use the normalized version of $|\psi^{[1]}\rangle$ as the input of the procedure. Repeat this until one exhausts all the eigenstates that have nonzero overlap in $|\psi_0\rangle$. For the remaining eigenstates having zero overlap with $|\psi_0\rangle$, we can generate another random state and remove the components of all previously found eigenstates and use the resultant state as the input. In this way, we can eventually exhaust all eigenstates. The benefit of this method is that the accuracy of each eigenstate is independent of one another.

\subsection{Simulation results: illustrative examples}

Let us illustrate the algorithm by considering the system to be five-level, i.e. qudit with $d=5$. We generate a $5\times5$ Hermitian matrix $H_5$, with $(H_5)_{ij}=(H_5)_{ij}^*=x+ y i$ and $x$ and $y$ uniformly sampled from the range $[-1,1]$ (except $y=0$ for the diagonal elements).  Here we only display its elements in the diagonals and below:
\begin{widetext}
\begin{equation}
\label{eq:H5}
H_5=\!\begin{pmatrix}
-0.0763231 & *& 
 * & * & 
 *\cr
 -0.51328\! +\! 0.0732759i & 
 0.691614& * & * &
 * \cr
  0.516039\! +\! 0.20004 i &  -0.884252\! -\! 
  0.248885i& -0.495554  & *  & *\cr
  -0.379429\! +\! 0.303255i &
 0.0981619 \!-\! 0.603679i & -0.484382\! -\! 0.134895i & 0.921927 &
 *\cr
 0.0142526\! +\! 0.421276i &
 0.635987\! +\! 0.0817911i & -0.450215\! -\! 0.808964i & 0.6387\! +\! 0.188711i &
 0.736562 
\end{pmatrix},
\end{equation}
whose eigenvalues $E_i$'s, sorted from the smallest to largest, are $\{-1.51593, -0.700576, 0.388005, 1.0888, 2.51793\}$. We also randomly generate a 5-component normalized vector to be the initial state,
\begin{equation}
\label{eq:psi0}
|\psi_0\rangle=(
0.506424, -0.370456 + 0.164849i, -0.444258 + 
  0.194814i, -0.0372888 - 0.33439i, -0.475495 - 0.0671035i)^T.
\end{equation}
The state $|\psi_0\rangle$ has an expected energy being $-0.525913$, with the probabilities $|\langle E_i|\psi_0\rangle|^2$  in the five eigenstates being, respectively, 
\begin{equation}\{0.554875, 0.0729256, 0.262368, 0.00841186, 0.10142\}.
\end{equation}
\end{widetext}

Next, we explore various combinations of $(\alpha,\beta)$ and $\Delta t$ in our classical simulations. Given that the optimal choice of $(\alpha,\beta)$ is such that $|\alpha\beta|=1/2$, i.e., within the one-parameter family $(1,e^{i\phi})/\sqrt{2}$, we first discuss the choice of the phase $\phi$ in this family. We have carried our a few simulations and displayed the results in Fig.~\ref{fig:QVNew}. 

\begin{figure*}
	(a)	\includegraphics[width=0.45\textwidth]{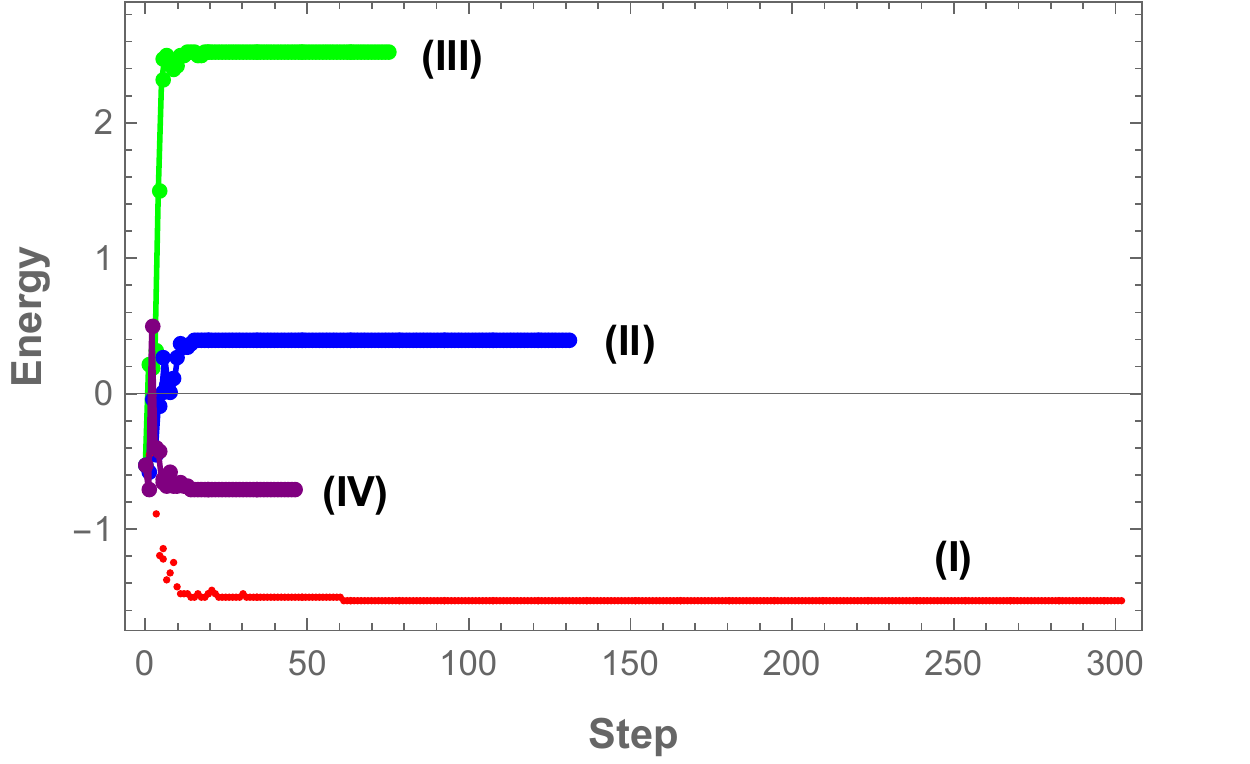}
	(b)	\includegraphics[width=0.47\textwidth]{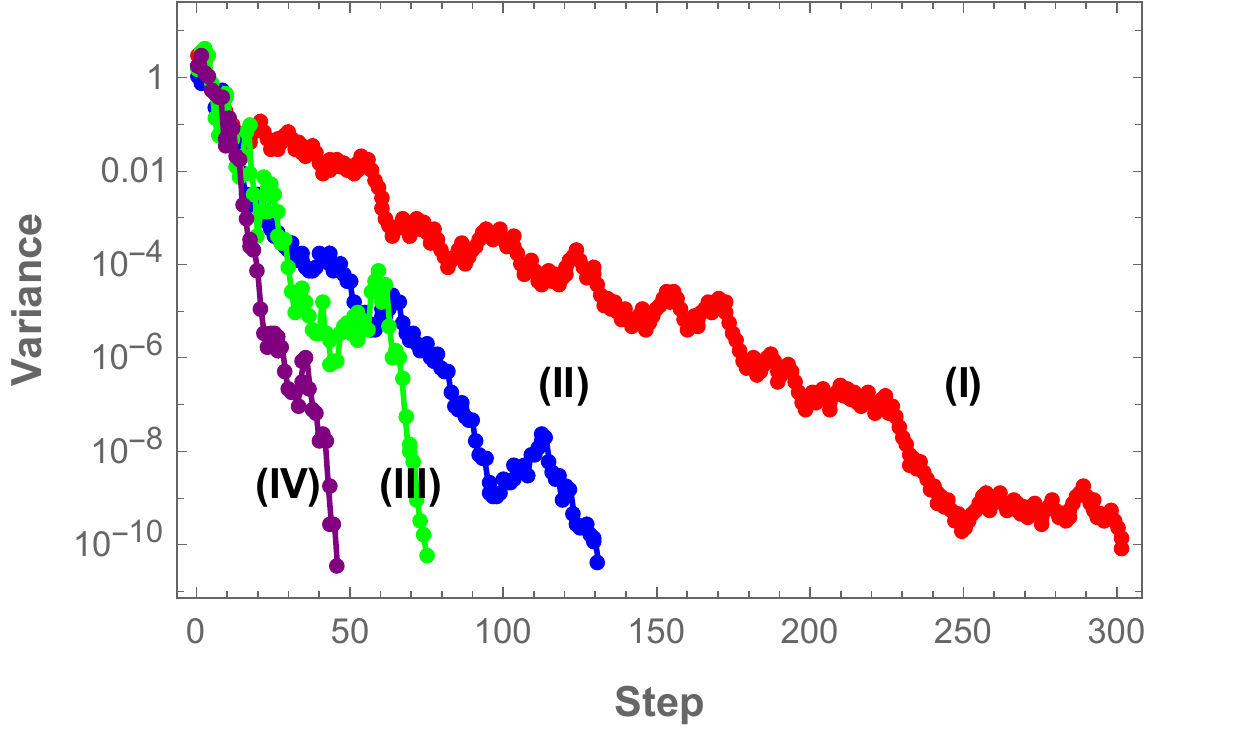}
		\caption{The iteration procedure using various choices of $\Delta t$ (arbitrary unit) and $\phi$ similar to those in Fig.~\ref{fig:QVNew}, except that $\alpha=\sqrt{3}/2$ and $\beta=1/2 \,e^{i\phi}$. See the main text for detailed discussions of the four types of choices (I) to (IV). In (a) the energy values (arbitrary unit) are recorded in the iteration; in (b) the energy variances (also in arbitrary units) are recorded.  
		 }
		\label{fig:QVNewA}
	\end{figure*}
	
\smallskip\noindent {\bf (I) Iterations with fixed $\Delta t$ and $\phi$}. With sufficient number of iterations,  even fixing $\Delta t =1.0$ and $\phi=0$ (i.e. the standard Hadamard test), eigenstates can be reached with increasing accuracy as the number of iterations increases. In the simulations, we terminate the iteration once the energy variance $\langle (\Delta h)^2\rangle$ has reached below $10^{-10}$. The specific example run takes as long as 365 steps and converges to the eigenenergy $E_3=0.388005$.

\smallskip\noindent {\bf (II) Iterations with fixed $\Delta t$ but $\phi$ from a given set}.
 Bying fixing $\Delta t=1.0$ but choosing $\phi$ from $k \pi/4$ (with $k=0,1,\dots,7$), the specific example run takes 67 steps to converge to the eigenenergy $E_2=-0.700576$.  

\smallskip\noindent {\bf (III) Iterations with fixed $\Delta t$ but random choice of $\phi\in[0,2\pi)$}. In the previous choice, $\phi$ is chosen from a set of values, here we consider choosing $\phi$ randomly from $[0,2\pi)$. In the example run, it takes 68 steps to converge to the eigenenergy $E_1=-1.51593$. 

\smallskip\noindent {\bf (IV) Iterations with varying $\Delta t$}. 
In the previous three cases, we do not need to change $\Delta t$. But by allowing $\Delta t$ to vary, the efficiency can be improved.  For example, by recycling $\Delta t$ from the set $\{100,100/3,100/3^2,100/3^3,100/3^4,100/3^5\}$ and using random $\phi$, it takes 43 steps to converge to the eigenenergy $E_5=2.51793$.

	We have also repeated the simulations using the above four types of choices but for $\alpha=\sqrt{3}/2$ and $\beta=e^{i\phi}/2$, and the results are shown in Fig.~\ref{fig:QVNewA}. We see that even without the optimal ancillary parameters, spectral projection can still be achieved. The steps it take to converge are not significantly larger than those using the optimal choice of the ancilla.

Let us compare our procedure to the QPE, in which the control-unitary needs to go as large power as $c-U^{2^{t_g-1}}$, in order to gain accuracy in $m$ binary digits, i.e. accurate up to $2^{-m}$, where $m=t_g-\log(2\epsilon+1/2\epsilon)$ and $1-\epsilon$ is the lower bound on the success probability of the QPE. To achieve the same accuracy as $2^{-33}\approx 10^{-10}$ in spectral projection by the QPE, one needs the number of ancillary qubits $t_g$ to be more than 33, and the power in $U$ differs in magnitude by $2^{33}$. In contrast, the ratio of the largest $\Delta t$ to the smallest used in our simulation (IV above) is only $3^5\approx 2^{8}$.  In the above (I)-(III), $\Delta t$ is fixed, but it takes more steps to converge to the desired accuracy.  

\begin{figure}
	(a)	\includegraphics[width=0.49\textwidth]{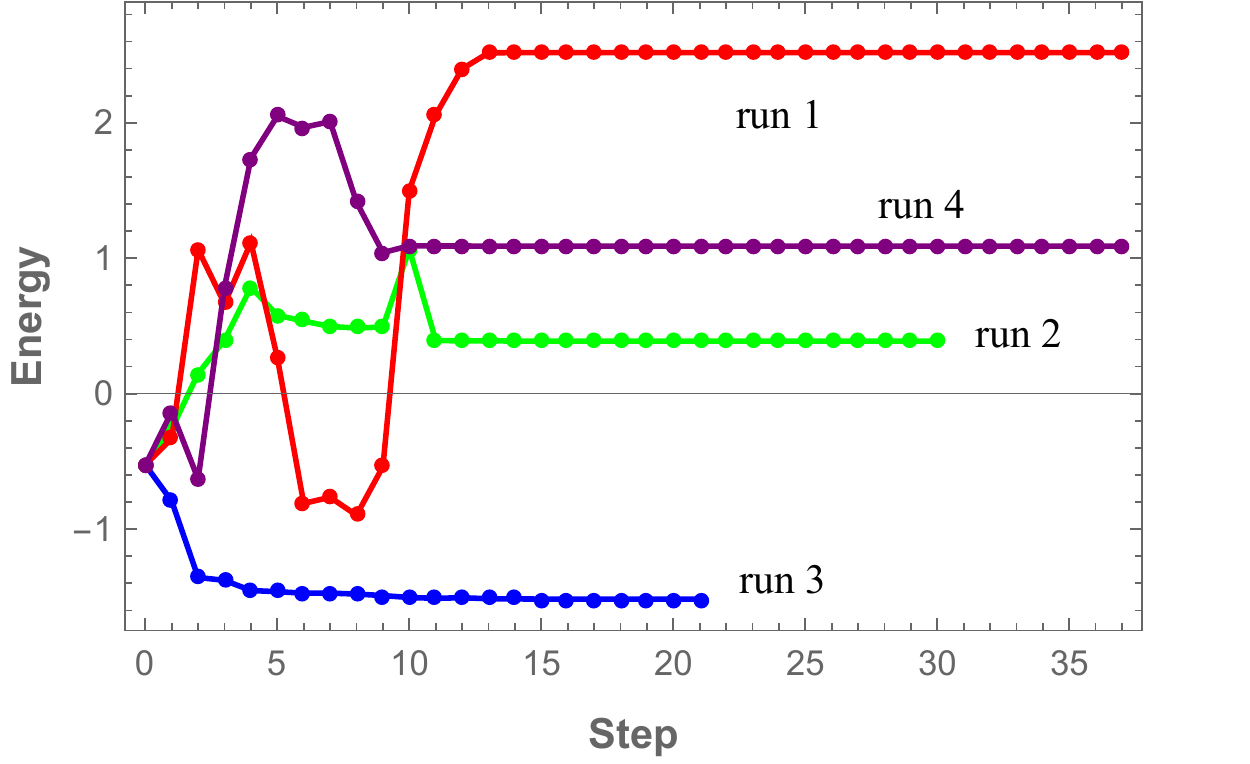}\\
	(b)	\includegraphics[width=0.49\textwidth]{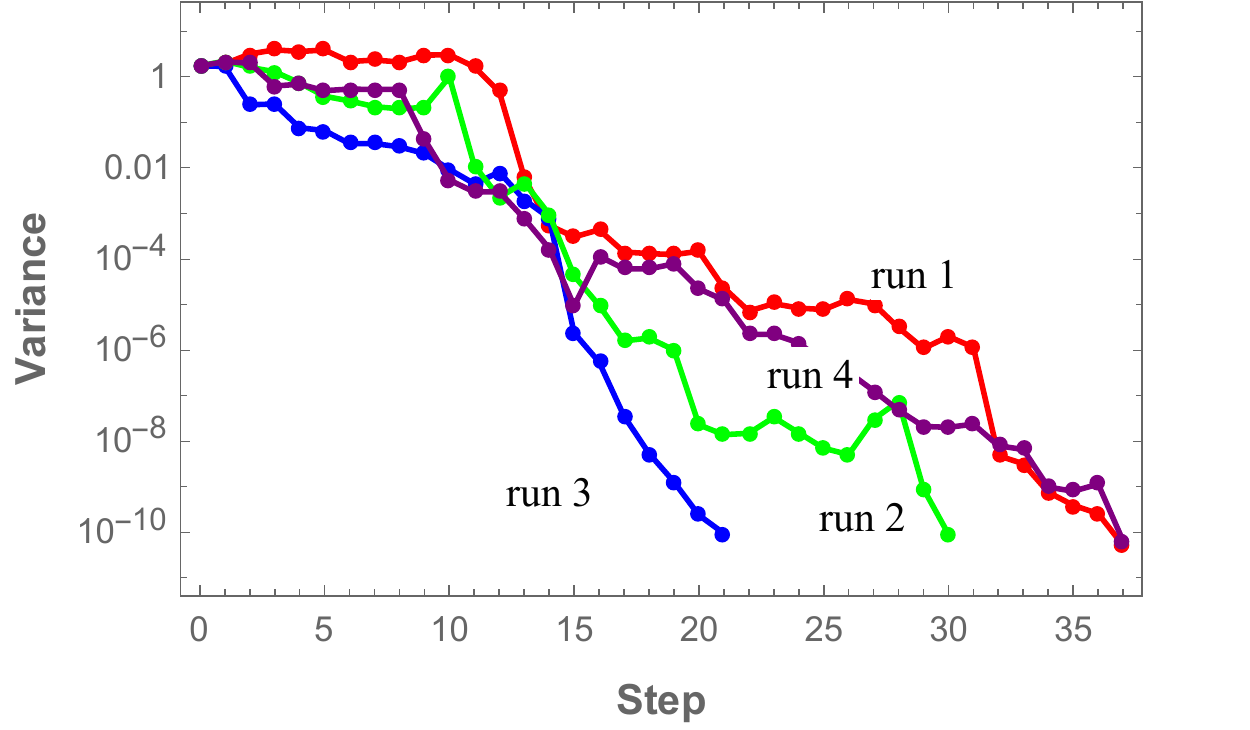}
		\caption{The iteration procedure using varying $\Delta t$ and random $\phi$. This corresponds to the choice (IV) in the main text. Here, $\Delta t$ is chosen from the set $\{100,100/3,100/3^2,100/3^3,100/3^4,100/3^5\}$, but perturbed by 1\% of random fluctuations: $\Delta t \rightarrow \Delta t (1+0.01x)$ with $x\in [0,1)$.  There are  four different runs but with the same initial state of the system. In (a) the energy values are recorded in the iteration; in (b) the energy variances are recorded.   Both the energy and its variance are displayed in arbitary units.
		 }
		\label{fig:QVNewPerturb}
	\end{figure}

\begin{figure}
		\includegraphics[width=0.48\textwidth]{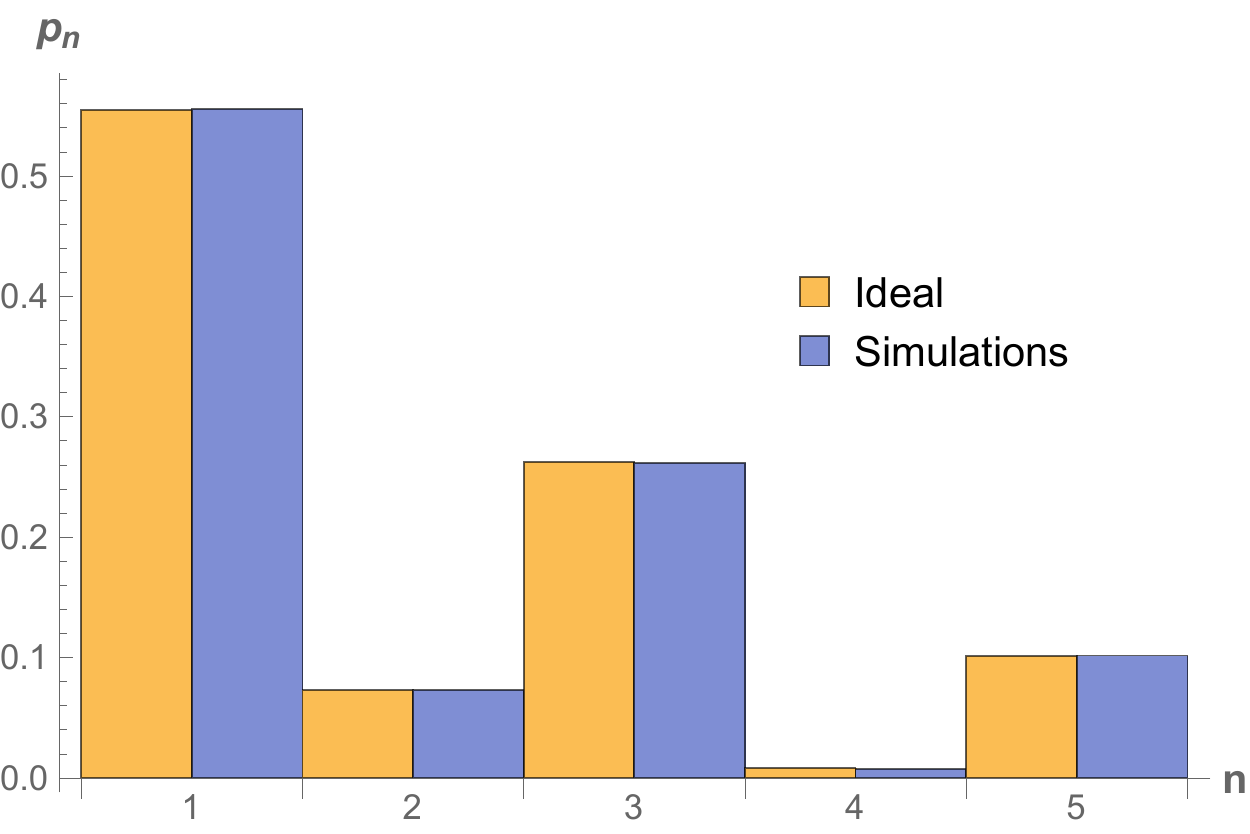}
		\caption{Eigenstates distribution $p_n$ after the procedure (simulations vs. ideal Born rule). The Born rule predicts that $p_n=|\langle E_n|\psi_0\rangle|^2$. The model under consideration is the $H_5$ Hamiltonian in Eq.~(\ref{eq:H5}) with the initial state in Eq.~(\ref{eq:psi0}). The horizontal axis $n$ labels the index of eigenstates with  eigenenergies ($E_n$) ordered from the lowest to the highest. 
	$\Delta t\in\{100,100/3,100/3^2,100/3^3,100/3^4,100/3^5\}$ and $\phi$ is chosen randomly each time in $[0,2\pi)$, and for each $\Delta t$ value we iterate 5 times.   Each run is terminated when $\langle (\Delta h)^2\rangle <10^{-10}$ and if more iterations are needed when all values in the $\Delta t$ list are used, we recycle the $\Delta t$ list from the beginning. 
		The statistics were obtaining by averaging over 10,000 runs. The final distribution obtained from the simulations is $\{0.5557, 0.0733, 0.2617, 0.0077, 0.1016\}$.
		}
		\label{fig:H5Chart}
	\end{figure}
In the QPE, the power of the unitary $U^{2^k}$ needs to be precise in order for the algorithm to work. The procedure that we propose here does not require precise $\Delta t$.  We have tested that the ability for the spectral projection does not depend on the precise values of  $\Delta t$ as above, and other sequences can be used. For example, a different sequence is used in Fig.~\ref{fig:QVNewPerturb} as an example by perturbing the previous set of $\Delta t$, and spectral projection is still achieved. 

In all of the above simulations, in addition to the energy value, the energy variance $\langle(\Delta h)^2\rangle$ is also recorded as the procedure is carried out. We have seen that on average, the energy variance indeed decreases.

\begin{figure*}
		\includegraphics[width=0.48\textwidth]{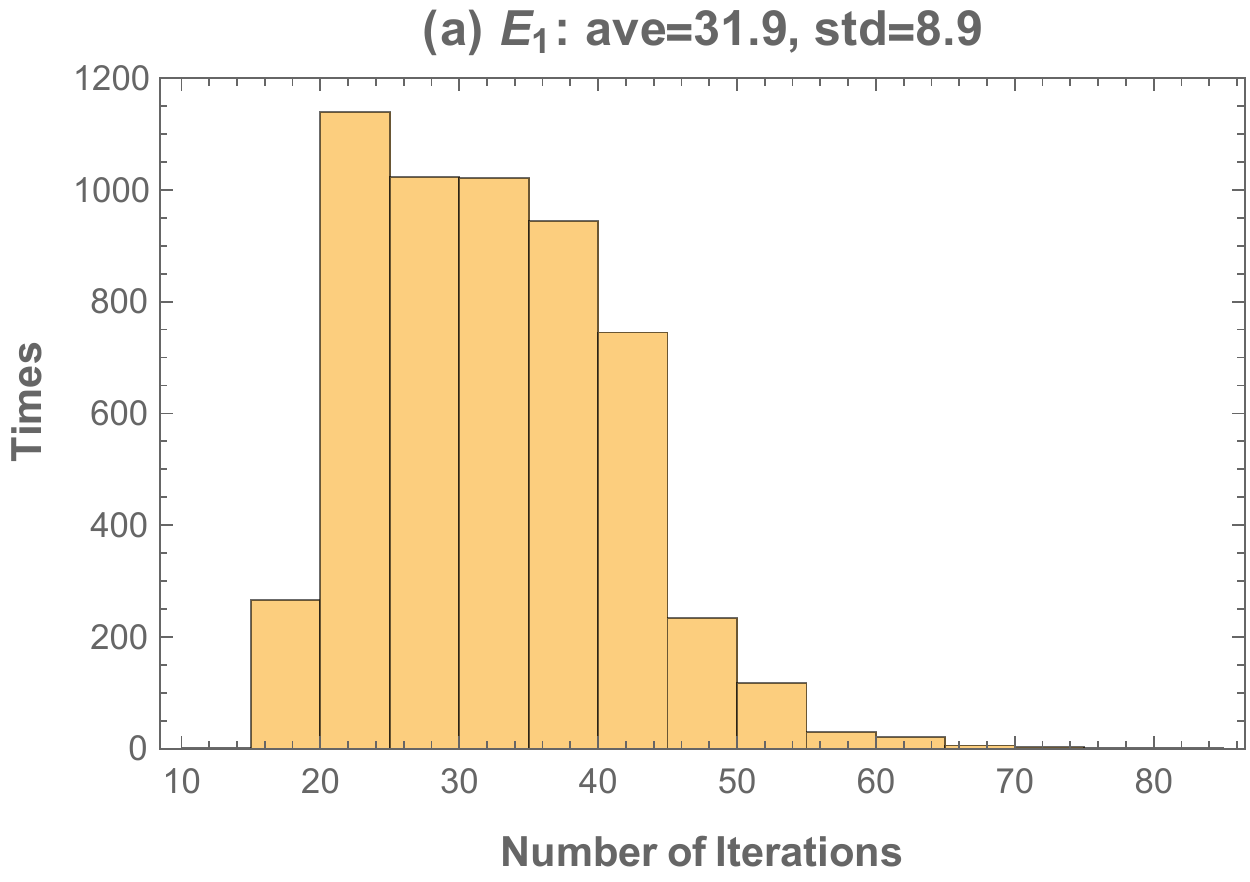}
		\includegraphics[width=0.48\textwidth]{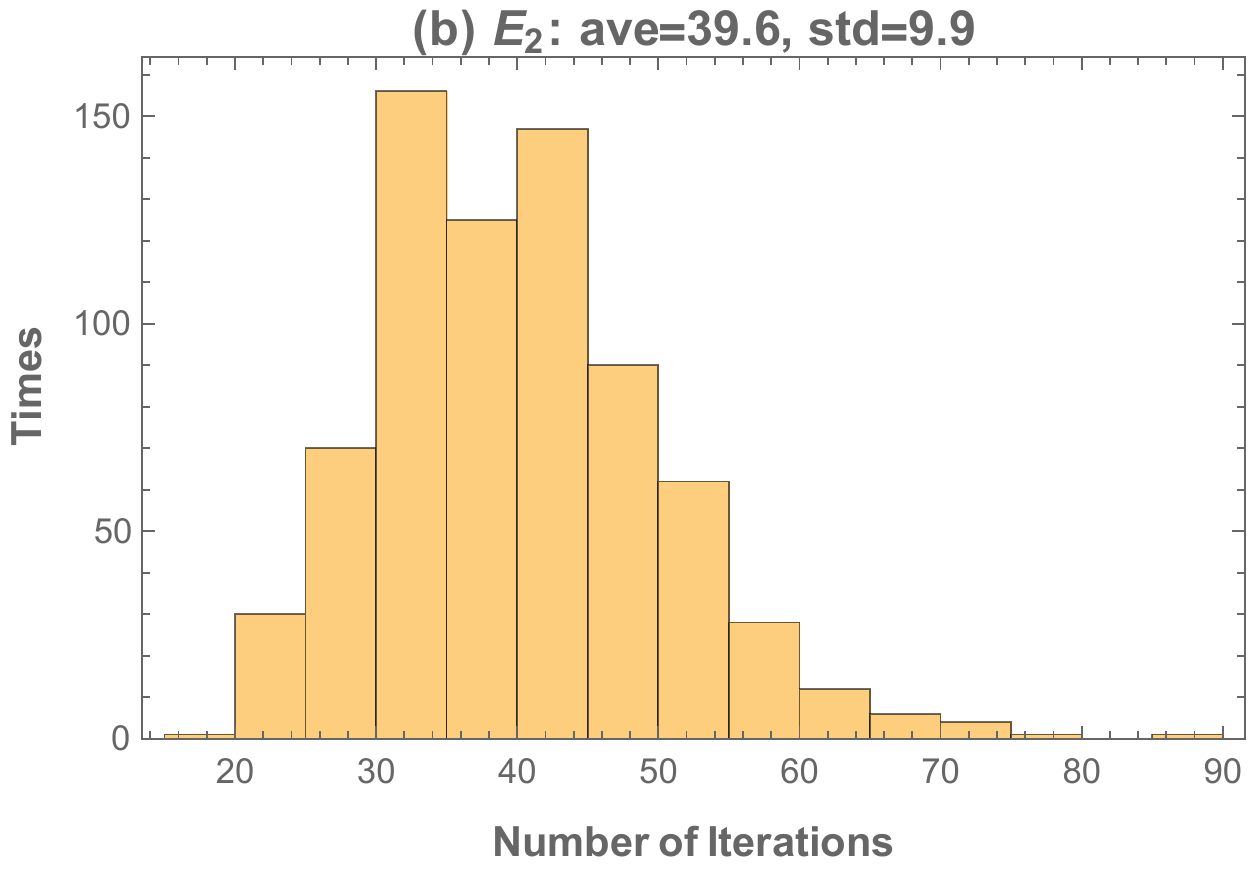}\\	
		\vspace{0.1cm}
		\includegraphics[width=0.48\textwidth]{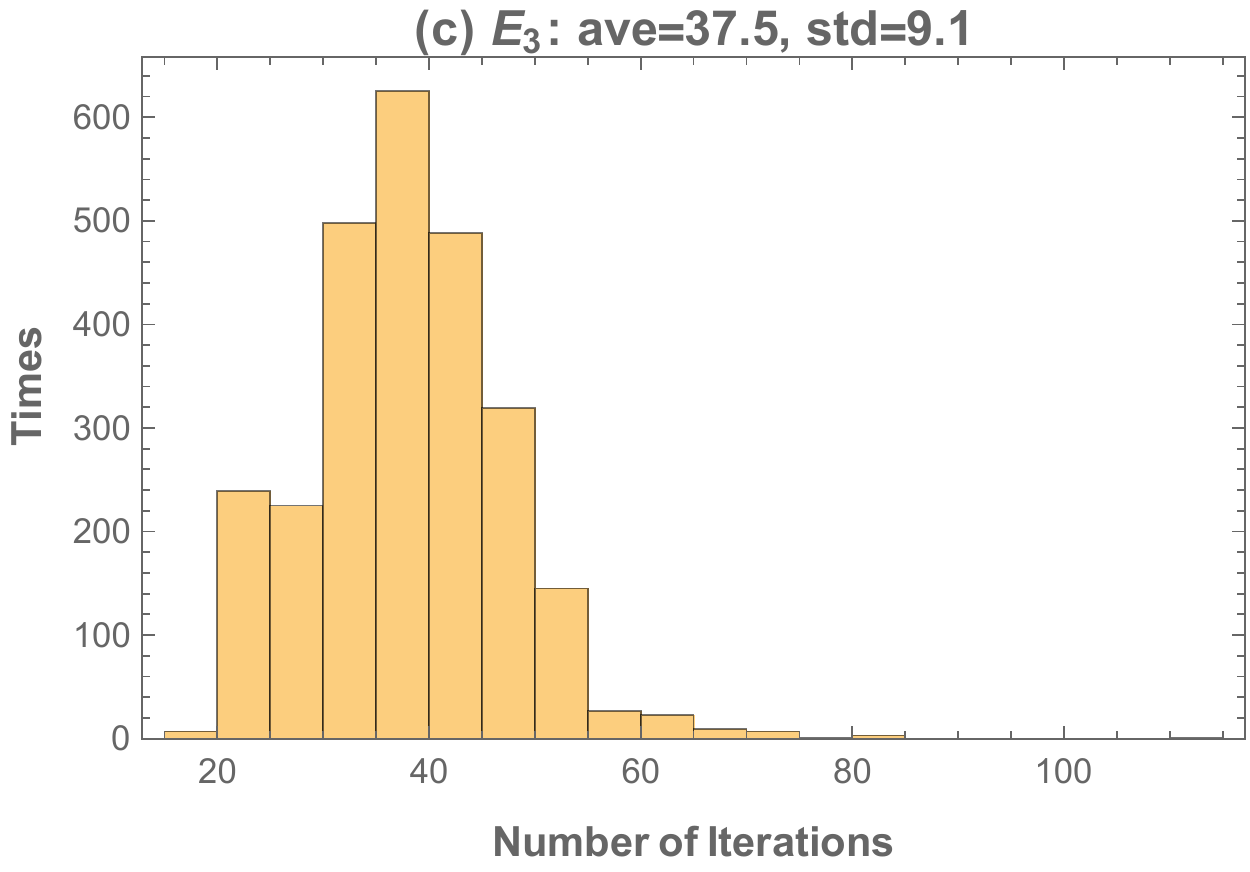}
	\includegraphics[width=0.48\textwidth]{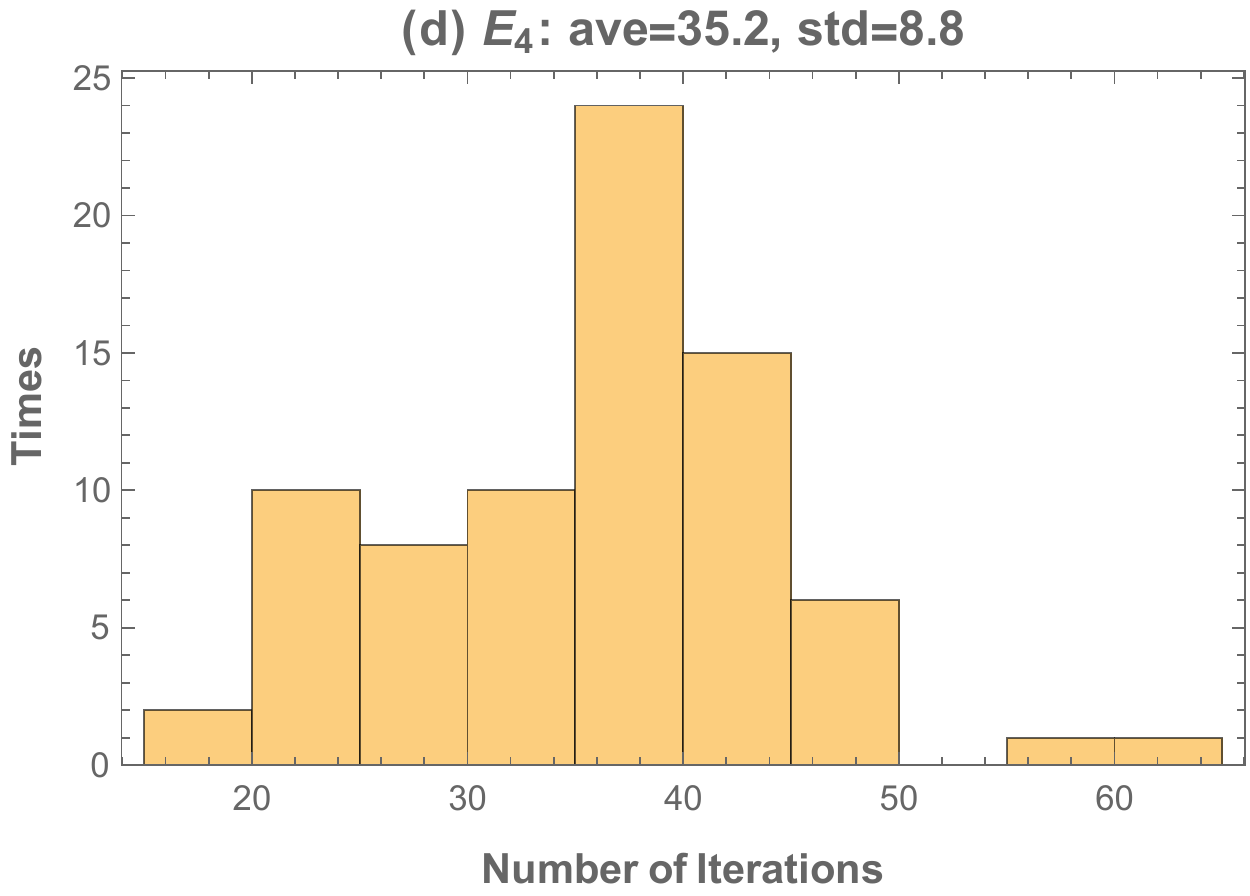}\\
		\vspace{0.1cm}
	\includegraphics[width=0.48\textwidth]{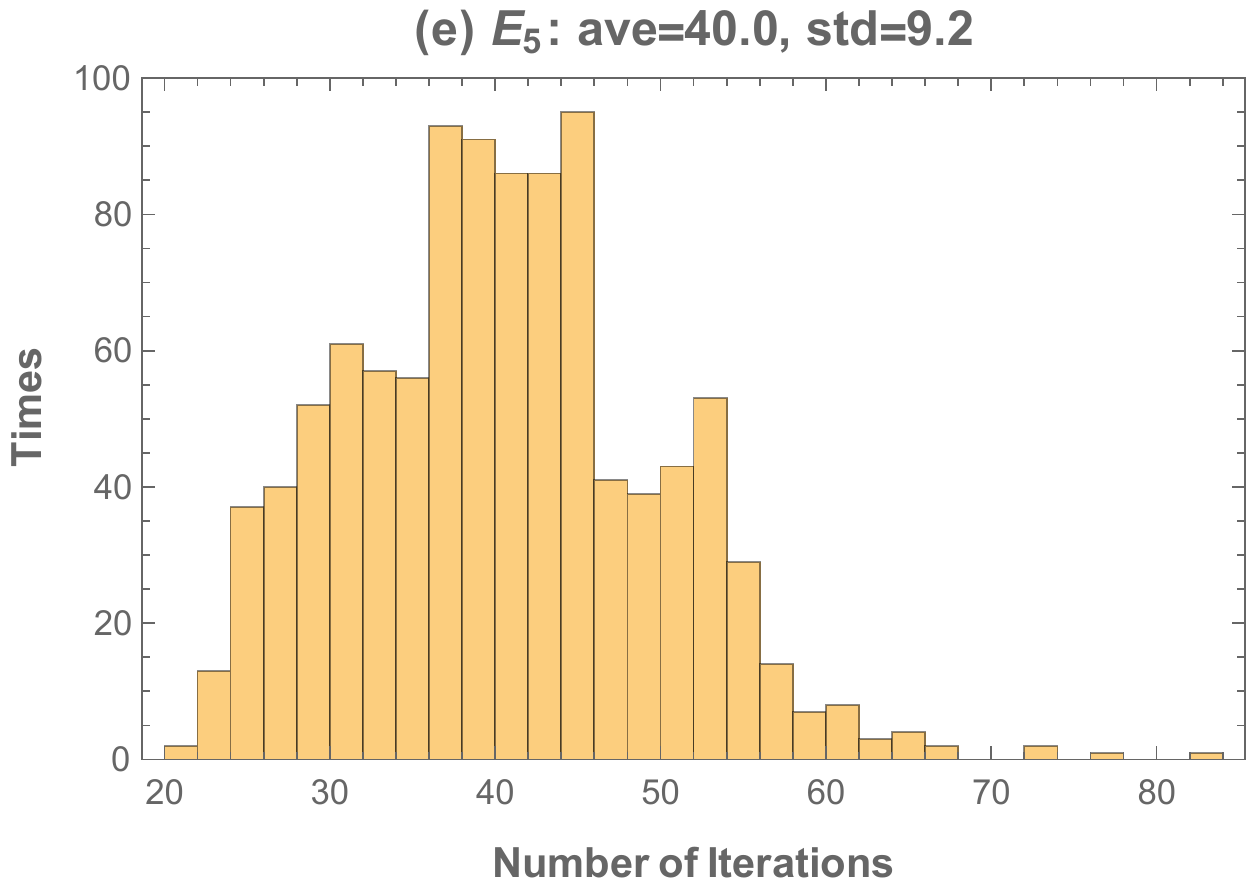}
		\caption{Histograms of the number of iterations to reach an eigenstate with an accuracy   $\langle (\Delta h)^2\rangle <10^{-10}$. 
		 }
		\label{fig:Histo}
	\end{figure*}

\subsection{Distribution of eigenstates: the Born rule}
Given that the iterations based on the primitive in Sec.~\ref{sec:prelim} lead to a procedure for projecting a system to eigenstates of an Hermitian operator $\hat{h}$, here, we investigate the distribution of eigenstates when this procedure is repeated many times. We again take the $H_5$ Hamiltonian~(\ref{eq:H5}) and the same initial state~(\ref{eq:psi0}) of the system and carry out simulations for our spectral projection algorithm.

As seen from the results in Fig.~\ref{fig:H5Chart} using 10,000 repetitions of the procedure, the distribution of the eigenstates agrees well with the Born rule, which predicts that the probability of the $n$-th eigenstate $|E_n\rangle$ is $p_n=|\langle E_n|\psi_0\rangle|^2$.  That  the Born rule applies can be explained as followed. Since the controlled unitary ${c-U}=|0\rangle\langle 0|\otimes\openone + |1\rangle\langle1|\otimes e^{-i\Delta t \hat{h}}$ commutes with the Hamiltonian $\hat{h}$ of the system, and hence with any eigenstate projector $|E_n\rangle\langle E_n|$. Therefore the expectation value of the observable  $|E_n\rangle\langle E_n|$ must be conserved and equals $|\langle E_n|\psi_0\rangle|^2$. Under the assumption and as observed above that the procedure leads to eigenstate projection, then the distribution $\{q_n\}$ of the projected eigenstates should remain the same as the initial distribution, i.e. $q_n=|\langle E_n|\psi_0\rangle|^2$.

We note that there is nothing special about the Hamiltonian $\hat{h}$, and our proposed algorithm works for any Hermitian operator. The Born rule will also apply. One may also regard our procedure as a method to realize  the statement in the measurement postulate.

\smallskip \noindent {\bf Number of iterations}. In addition to the Born rule, we also investigate how many iterations are needed to reach a desired accuracy, e.g.  $\langle (\Delta h)^2\rangle <10^{-10}$.  In the same simulation for the study of the Born rule above, we also keep track of the number of iterations in each run it takes to reach that accuracy. The results are shown in Fig.~\ref{fig:Histo} using  histograms. As observed, the number of required iterations is not narrowly peaked  and  this reflects the randomness in the ancilla measurement outcome and the state dependence in the outcome probability. 

	\section{Spectral projection algorithm applied to the transverse-field Ising model}
	\label{sec:physical}

	\begin{figure*}
(a)		\includegraphics[width=0.45\textwidth]{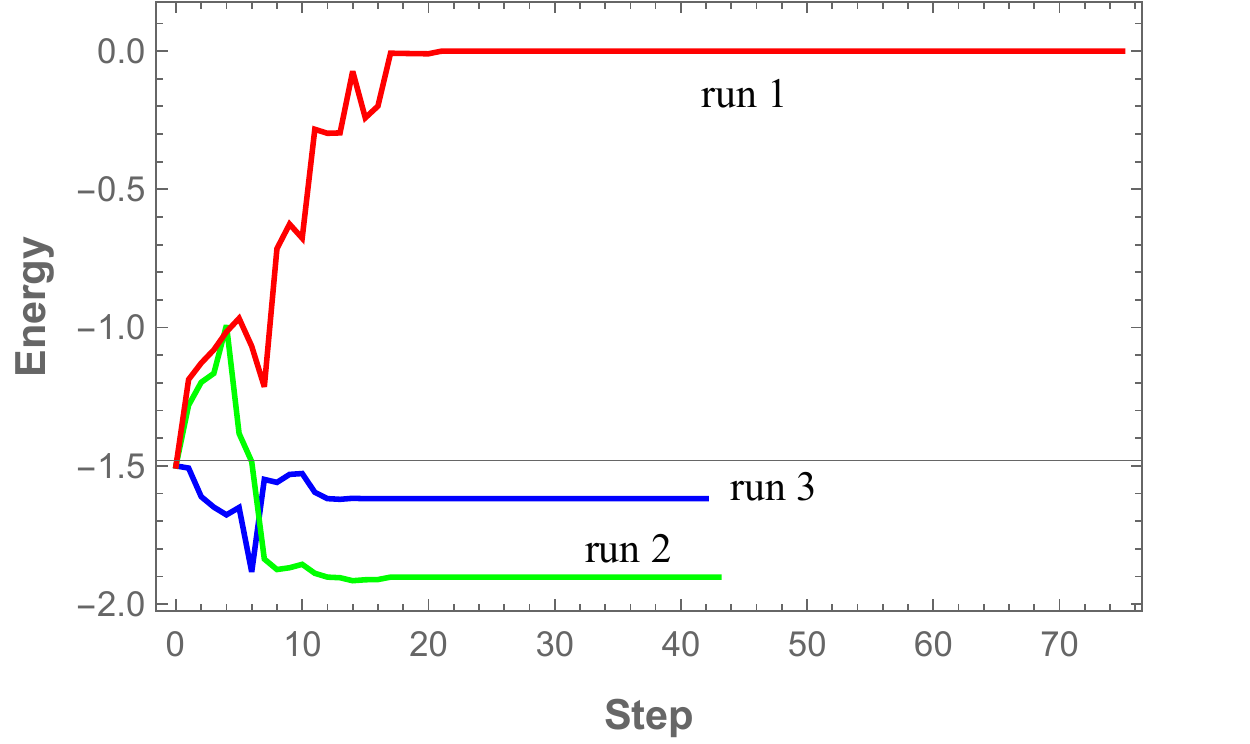}
(b)		\includegraphics[width=0.45\textwidth]{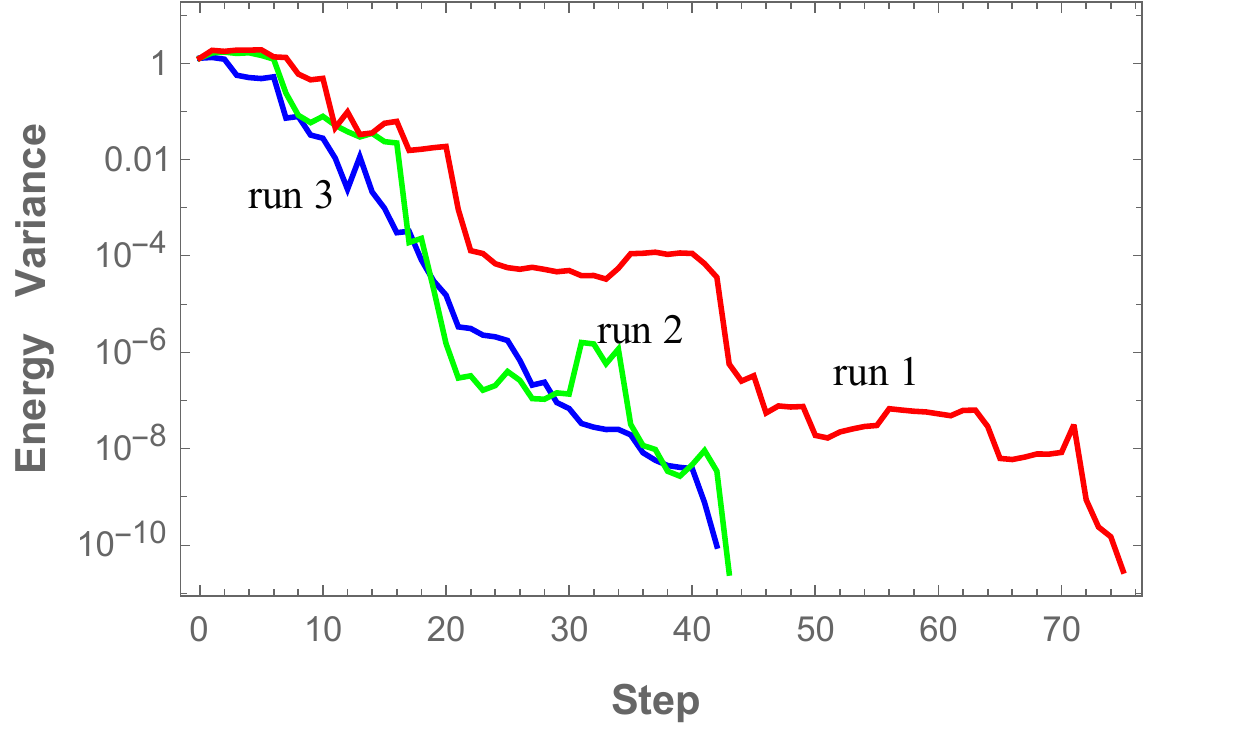}\\
(c)		\includegraphics[width=0.90\textwidth]{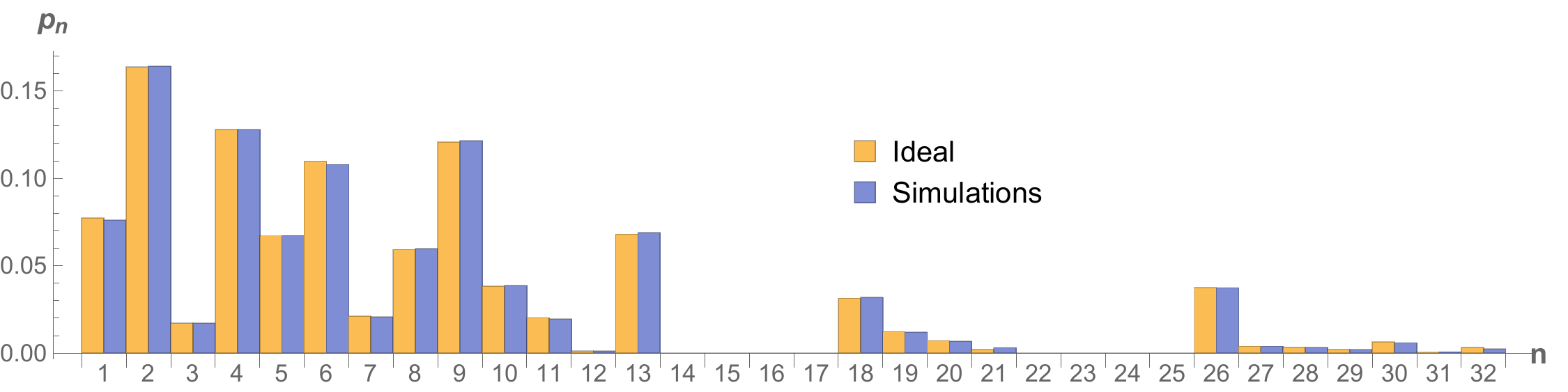}
		 \caption{Example simulations on spectral projection for 5-qubit transverse field Ising model. (a) and (b) show the traces of energy and its variance in arbitrary units, respectively. There are three different runs (but with the same initial state of the system). Each run is terminated when $\langle (\Delta h)^2\rangle<10^{-10}$. The phase parameter $\phi$ in the ancilla state is chosen randomly at each step  and $\Delta t$ is chosen from the set $\{100,100/3,100/3^2,100/3^3,100/3^4,100/3^5\}$ and each $\Delta t$ repeated 5 times.  The iterations continue by recycling the $\Delta t$ set until the desired precision is met. (c) The bottom panel compares the distribution of projected eigenstates in 10,000 simulation runs with the ideal Born rule.}
		\label{fig:Ising}
	\end{figure*}
Here we consider physical models, such as the 	Ising model in a transverse field (with the periodic boundary condition), 
\be
\label{eqn:Ising}
{H}_{\rm TFI}(g)= \sum_{i=1}^{N_q} \left[g\,\sigma^x_{i}\sigma^x_{i+1} -(1-g)\sigma^z_i\right].
\ee
Our parameterization is slightly different from that in the literature. The spin-spin coupling strength is $J=g$ (antiferromagnetic if $J>0$) and the external field is $B=(1-g)$.  
In Fig.~\ref{fig:Ising}, we take $g=0.5$ (the critical point in the large $N_q$ limit), $N_q=5$ and the initial state $|\psi_0\rangle=|+-+-+\rangle$ and simulate the spectral projection procedure. The values of $\phi$ are randomly chosen and those of $\Delta t$ are listed in the caption. We see in Fig.~\ref{fig:Ising} that spectral projection can be achieved with accuracy of $10^{-10}$ by using $\Delta t$ that ranges less than three orders of magnitude. To use the QPE for spectral projection, it will require the unitary controlled by the ancilla to raise to at least $2^{32}$, which is far from practical at present.

We also compare the distribution of projected eigenstates in the simulation with the ideal Born rule. In the case of degeneracy, we assign the portion according to the overlap square with these degenerate eigenstates. This again confirms the Born rule of our spectral projection procedure in a spin model.

\smallskip\noindent {\bf The use of the ancillary state $|A\rangle=|\pm\rangle$}. In our simulations for the Ising model we have encountered cases where the use of $|\pm\rangle$ in the ancillary state has caused the system to flow to certain class of states which under further iterations do not change the energy, despite that they were not eigenstates. But we have not observed such phenomena in the random Hamiltonian case explored earlier. This can be explained by the expressions in Eq.~(\ref{eq:dE2+}), which shows that when $\langle\hat{h}^3\rangle -\langle\hat{h}^2\rangle \langle\hat{h}\rangle=0$, the energy does not change to lowest order in $\Delta t$. This occurs when 
\begin{equation}
\label{eq:fixedpoint}
|\psi\rangle=\sum_{i; E_i=-E} a_{i} |E_i\rangle +\sum_{j; E_j=E}b_j|E_j\rangle,
\end{equation}
 as one can verify that
\begin{eqnarray}
\langle\hat{h}\rangle&=&\left(\sum_{j;E_j=E} |b_j|^2-\sum_{i;E_i=-E} |a_i|^2\right) E,\\
\langle\hat{h}^2\rangle&=&E^2,\\
\langle\hat{h}^3\rangle&=&\left(\sum_{j;E_j=E} |b_j|^2-\sum_{i;E_i=-E} |a_i|^2\right) E^3,
\end{eqnarray}
and, hence, the above condition is satisfied.  The state does change under the iteration, but not the magnitudes $|a_i|^2$ and $|b_j|^2$ (after proper normalization). In the case of the transverse-field Ising model, there are eigenstates of opposite energies, and thus it can happen the system is driven to states of the form~(\ref{eq:fixedpoint}). In our example from random Hermitian matrices, there are no  eigenstates of opposite energies.

\section{Effect of decoherence}
\label{sec:decoherence}
\begin{figure}
		\includegraphics[width=0.48\textwidth]{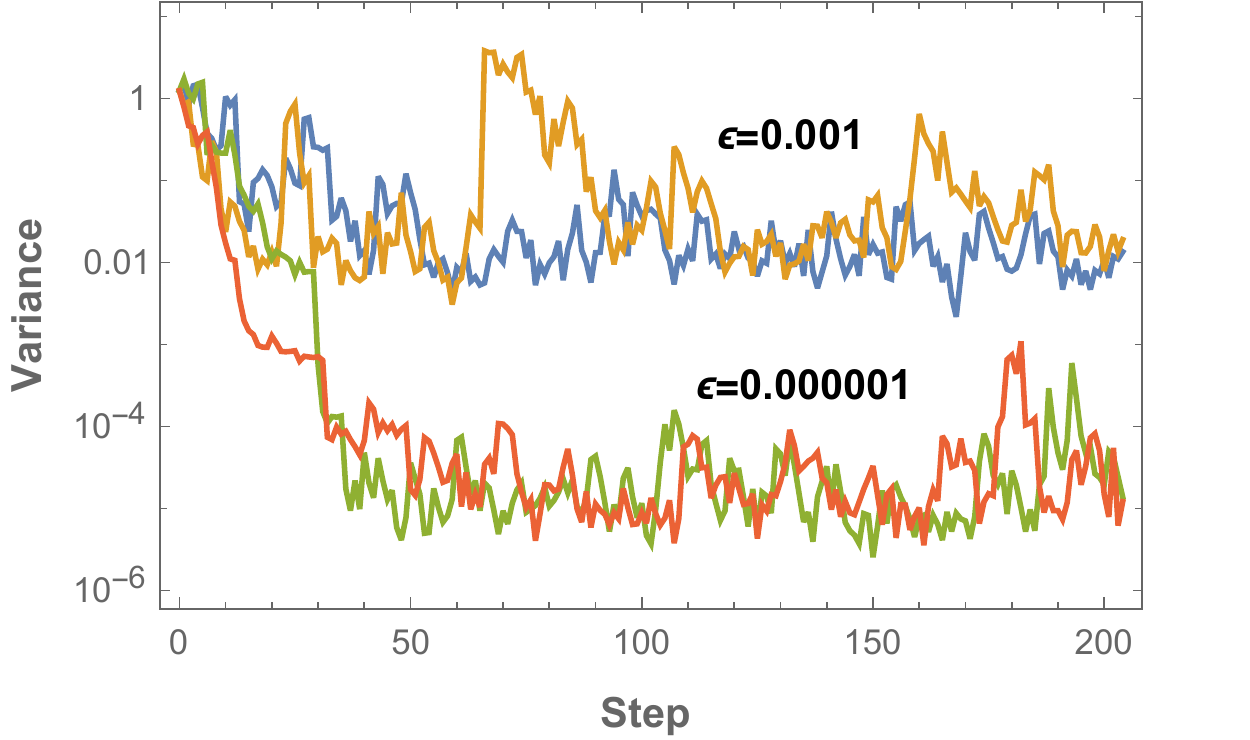}
		 \caption{Energy variance (in an arbitrary unit) in presence of decoherence. We apply the depolarizing channel at each step of the iteration, with two different $\epsilon=10^{-3}$ and $10^{-6}$. Two runs are performed for each respective $\epsilon$. The Hamiltonian is the transverse-field Ising model with $g=0.5$. It is seen that the energy variance is larger than $5\epsilon$. There are 204 steps in each run.}
		\label{fig:DecohEV}
	\end{figure}
Our method in general does not protect against decoherence.  Let us consider a simple depolarizing channel  $D(\rho_i)=(1-\epsilon)\rho_i +\epsilon \openone_i/2$ apply to every system qubit. Here  we assume the ancillary control qubit is relatively error free, and this reminds us of the assumption in the so-called DQC1 quantum computing model~\cite{Knill1998}, where only one qubit is clean. Due to the depolarizing channel, the state remains in the original un-decohered state with a probability approximately $(1-\epsilon)^{N_q}$, where $N_q$ is the number of qubits in the system, but the remaining portion $1-(1-\epsilon)^{N_q}$ can contribute substantially to the energy change and its variance.  If the decoherence is applied at each step, then our procedure will have an error of order at least $1-(1-\epsilon)^{N_q}\approx N_q \epsilon$ for small $\epsilon$ at each step of the iteration.  This is confirmed in our numerical simulations, as shown by the record of energy variance in Fig.~\ref{fig:DecohEV}.

\begin{figure}
	(a)\\	\includegraphics[width=0.49\textwidth]{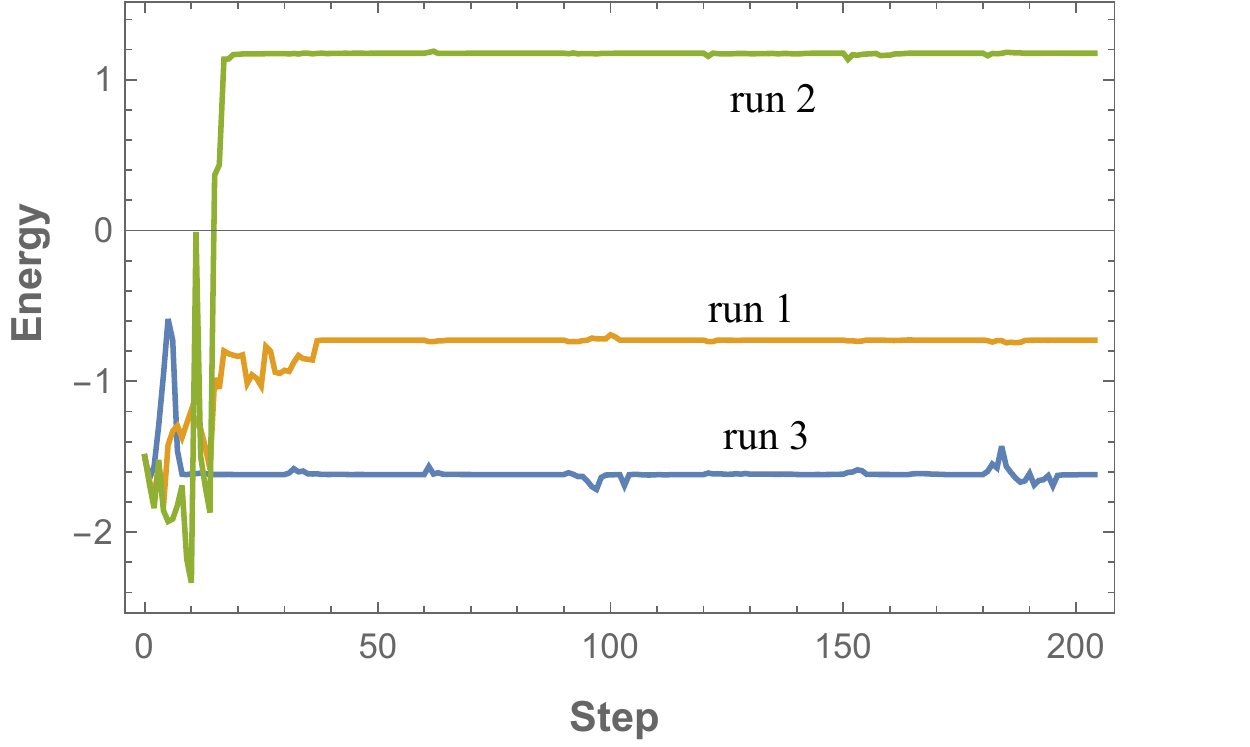}\\
	(b)\\	\includegraphics[width=0.49\textwidth]{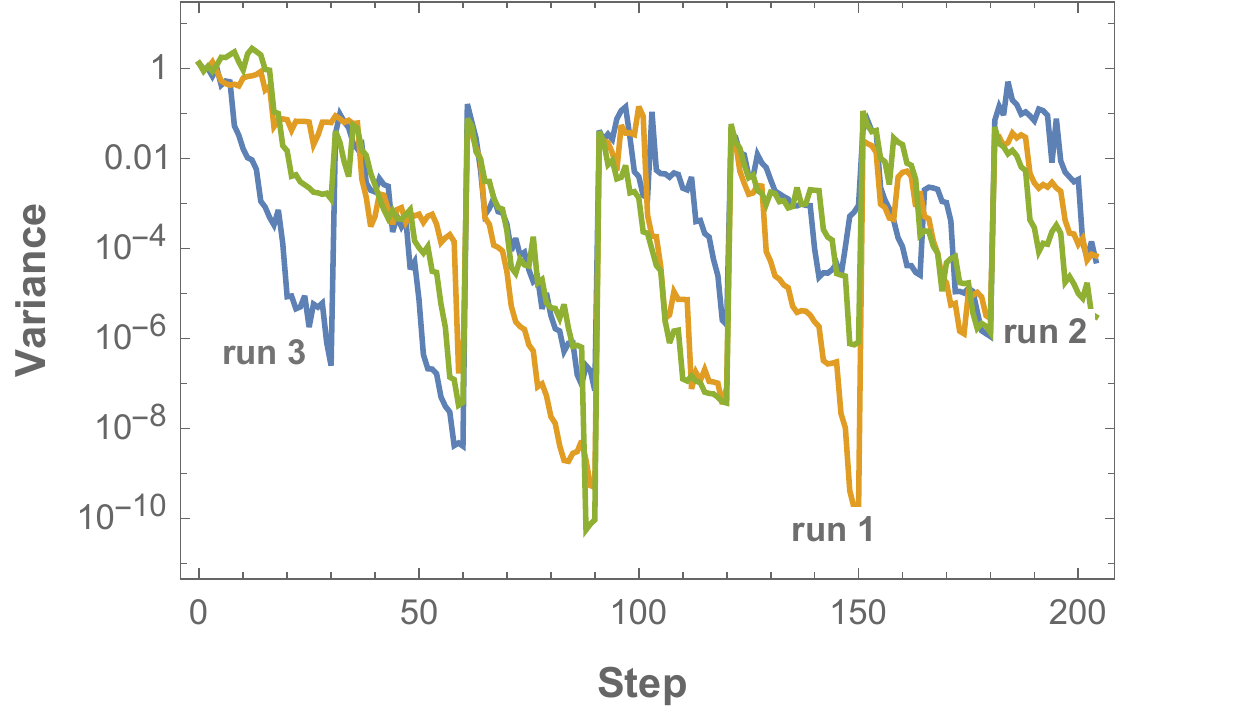}
		\caption{Energy (top) and energy variance (bottom), similar to the simulations in Fig.~\ref{fig:DecohEV}, except that $\epsilon=0.01$ and the depolarizing channel applies only every 30 steps, starting at step 1 and ending at step 181. There are 204 steps in each run of the three runs. Both the energy and its variance are displayed in arbitary units.
		 }
		\label{fig:Decoh}
	\end{figure}

However, we imagine a contrived scenario that the depolarizing channel acts only, e.g., every 30 steps. Then the procedure can achieve better accuracy in between two strikes of the decoherence. This is illustrated in Fig.~\ref{fig:Decoh}. The `disruptions' due to decoherence are visible, especially in terms of the upward jump in the energy variance in Fig.~\ref{fig:Decoh}b. If there can be sufficient number of iteration steps carried out before decoherence takes place, then the system can converge close to an eigenstate. Of course, the depolarizing channel takes the system out of the eigenstate and the subsequent iterative spectral projection procedure may take the system towards another eigenstate.
Since the decoherence process does not commute with the system's Hamiltonian, our procedure in the presence of decoherence may serve at best a robust way of finding arbitrary eigenstates, rather than a robust way of spectral projection.

\begin{figure}[t]
	(a)\\	\includegraphics[width=0.48\textwidth]{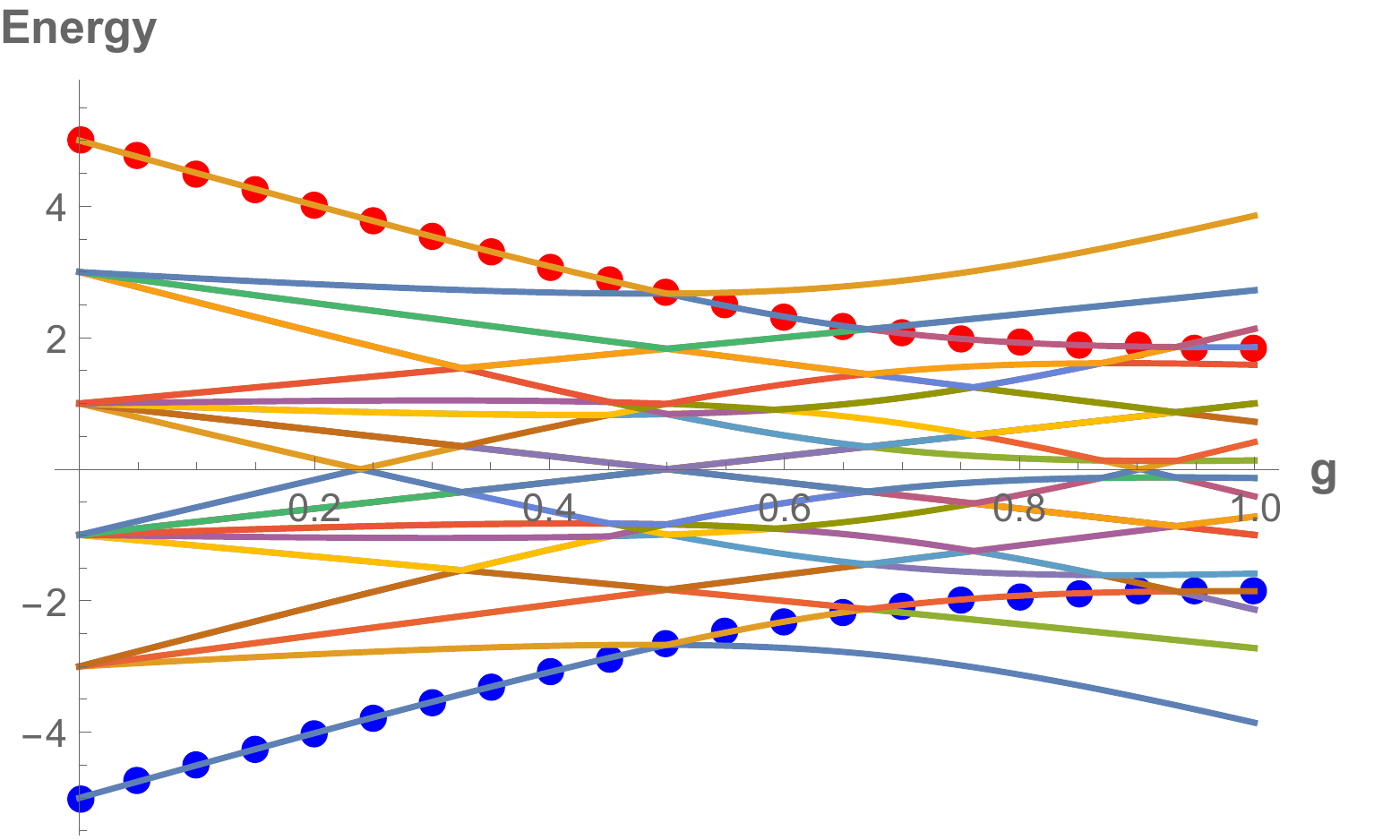}\\
	(b)\\		\includegraphics[width=0.48\textwidth]{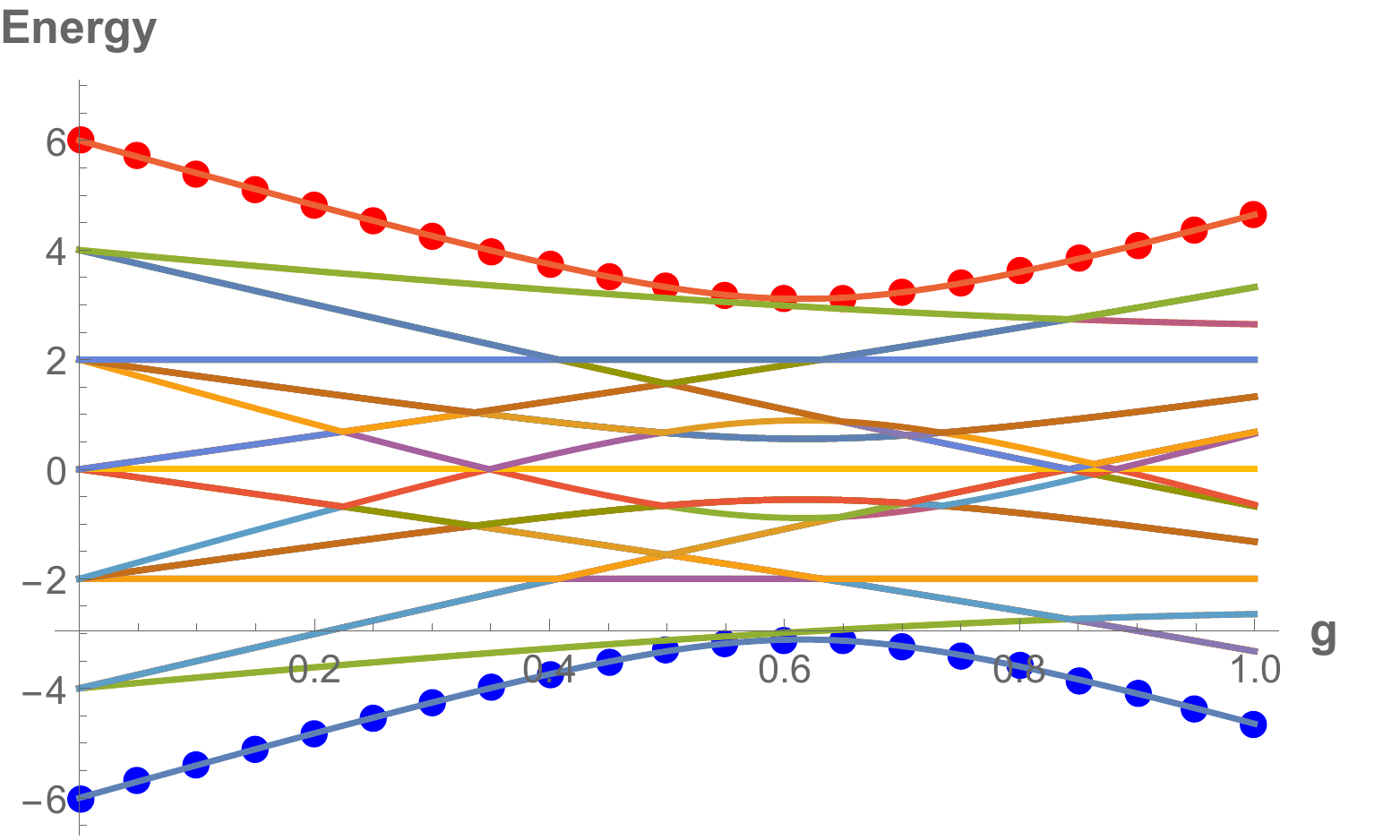}
		\caption{Application of our spectral projection algorithm in the quantum annealing as the subroutine. The figures display the energy (in an arbitrary unit) after projecting to the eigenstates vs. $g$ for the  transverse-field XzY model at $r=0.5$, i.e. $H_{\rm XzY}(g=0,r=0.5)$ for (a) $N=5$ qubits, and (b) $N=6$ qubits. The curves represent eigenenergies as a function of $g$. The procedure starts with two different initial states: (1) [(blue) dots that start on the lowest curve] the  ground state $|00000\rangle$ of $H_{\rm XzY}(g=0,r=0.5)$, and (2) [(red) dots that start on the top curve] the highest-energy state $|11111\rangle$ of $H_{\rm XzY}(g=0,r=0.5)$. All the energy levels of $H_{\rm XzY}(g,r=0.5)$ are also shown by solid curves. (a) Due to energy level crossings, the ground state transits to a higher excited state after the crossing, and the highest energy state transits to a lower energy state after an associated crossing. Quantum annealing does not work if there is any level crossing. 
		(b) Due the existence of a respective small gap,  the initial ground state ends up at the final ground state and the initial highest-energy state ends up also at the final highest-energy state. }
		\label{fig:gHXnY}
	\end{figure}

\section{Spectral projection as a subroutine in the quantum annealing algorithm}
\label{sec:subroutine}
We begin by describing the idea of quantum annealing and related algorithms. One of the first proposed quantum annealing methods is to use imaginary-time Schr\"odinger's equation proposed by Finnila et al.~\cite{FirstQA}.  The one that is close to the modern AQC~\cite{Adia1,Adia2} is proposed by Kadowaki and Nishimori~\cite{KadowakiNishimori}, where the Hamiltonian is the combination of the time-independent Ising model and a time-dependent transverse field.  The evolution of the quantum state was discussed in terms of real-time Schr\"odinger's equation  that takes the system in the ground state of the large-field limit towards that of the zero-field limit. The AQC similarly has a Hamiltonian $H(g)$ that interpolates between a simple Hamiltonian $H(g=0)$ with an easily prepared ground state  $|G(0)\rangle$ and the final Hamiltonian $H(g=1)$ that encodes the solution of certain problem in the ground state of $H(g=1)$. Provided the minimum gap of $H(g)$ is not too small, then evolving under the Hamiltonian via a suitable path $g(t)$ will take the initial ground state very close to the final ground state at the end of the evolution,
\begin{equation}
|\Psi(T)\rangle =\hat{T} e^{-i\int_{0}^{T}  H(g(t)) dt}|G(0)\rangle,
\end{equation}
  where $\hat{T}$ indicates that the integration is time-ordered, and $T$ is the total time duration.
  
  The key idea of the QSA by Somma et al.~\cite{Somma2008} is to exploit the quantum Zeno effect and replace the unitary evolution by measurement in the eigenbasis of $H(g_i)$, in a successive sequence of discrete $g_i$ ($0<g_1<g_2<\dots <g_T=1$). If the overlap of successive ground state $|\langle G(g_k)|G(g_{k+1})\rangle|\ge 1-\mu^2$ is sufficiently close to unity, then by the quantum Zeno effect, the final state after the whole sequence of measurement should be very close to the final ground state $|G(g=1)\rangle$. The standard QPE and a randomization procedure were proposed in Ref.~\cite{Somma2008} to achieve the measurement approximately. Below, we use our spectral projection algorithm for the measurement in the QSA and perform classical simulations for two different Hamiltonians, and we loosely refer to this also as quantum annealing.
\subsection{Transverse-field XzY model}
Here, we consider a different spin chain~\cite{DegerWei2019} than the Ising model:
\begin{eqnarray}
\label{eqn:XzY}
{H}_{\rm XzY}(g,r)&=& \sum_{i=1}^{N_q} \left[-g\left(\frac{1+r}{2}\sigma^x_{i-1}\sigma^z_{i}\sigma^x_{i+1} +\right.\right.\\
&&\left.\left.\frac{1-r}{2}\sigma^y_{i-1}\sigma^z_{i}\sigma^y_{i+1} \right)-(1-g)\sigma^z_i\right].\nonumber
\end{eqnarray}
One reason of choosing this transverse-field XzY model is because,  for the qubit number $N_q$ being odd, there is a crossing in the lowest few energy levels when the parameter $g$ is varied; see e.g. Fig~\ref{fig:gHXnY}. But for $N_q$ being even, there is a small gap above the ground state (for finite $N_q$). Therefore, it is interesting to compare the two different cases (but in the same model) for the quantum annealing. In our simulations, we will take $r=0.5$.

We begin with the initial state either as the ground state or the highest-energy state of $H_{\rm XzY}(g=0,r=0.5)$ and run the simulations for the quantum annealing with our spectral projection algorithm as a subroutine. The projection subroutine works by performing 180 times the primitive in Sec.~\ref{sec:prelim}, thereby approximately projecting the system to eigenstates of $H_{\rm XzY}(j\Delta g,r=0.5)$, where in this simulation $\Delta g=0.05$ and $j$ successively goes from 1 to 20, reaching $g=1$ at the end. We see, in Fig.~\ref{fig:gHXnY}a with $N_q=5$, that the quantum annealing does not work as there is an energy level crossing and the state of the system follows its path smoothly in the energy space crossing the lowest energy curve, and similarly for the initial highest-energy case. However, the quantum annealing indeed does work when there is a gap throughout the range of $g$ (except at the end) for the $N_q=6$-qubit case in Fig.~\ref{fig:gHXnY}b.

\subsection{Transverse-field Ising model}
\begin{figure}[t]
	(a)\\	\includegraphics[width=0.48\textwidth]{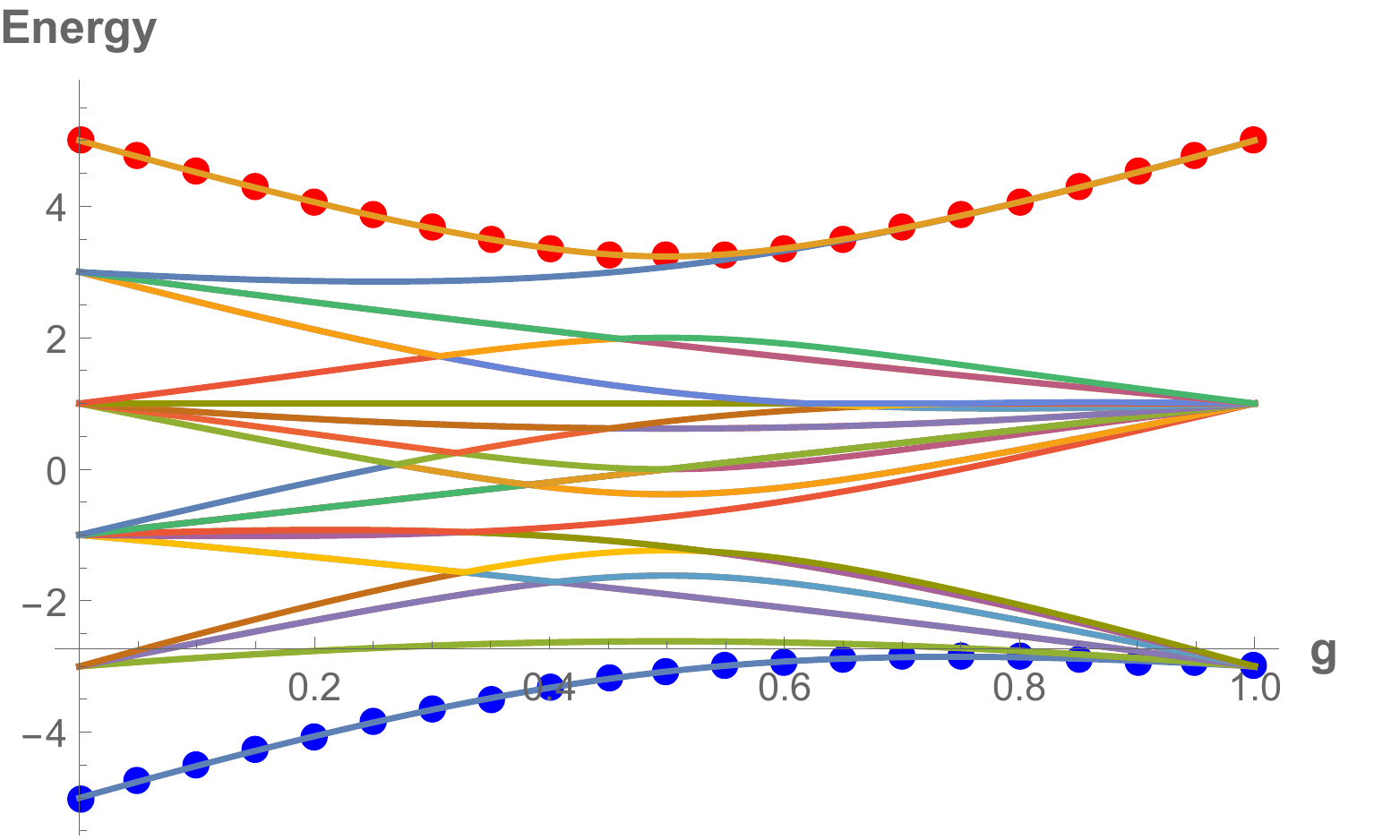}\\
	(b)\\	\includegraphics[width=0.48\textwidth]{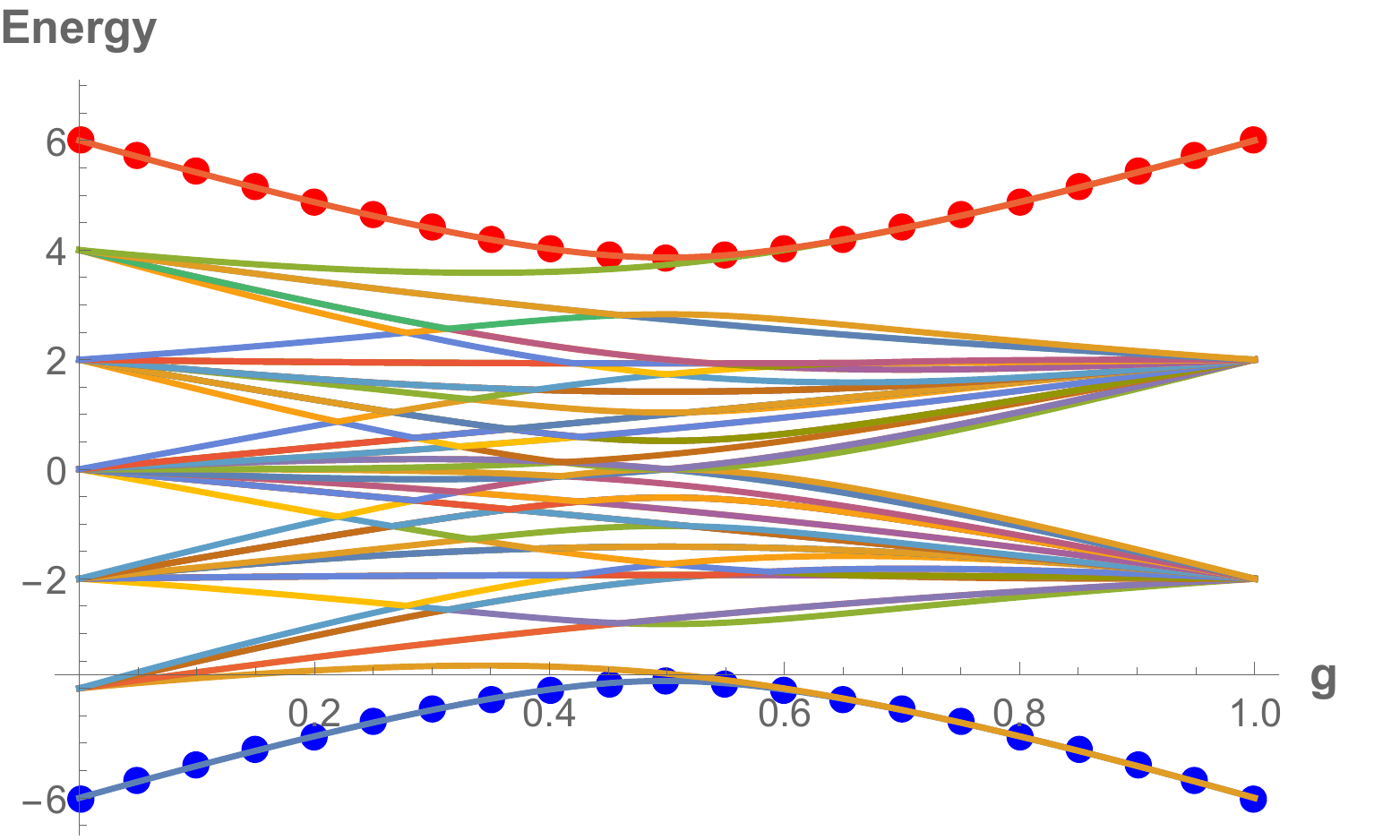}
		\caption{Application of our spectral projection algorithm in the quantum annealing as the subroutine for the transverse-field Ising model ${H}_{\rm TFI}(g)$. The figure shows the energy (in an arbitrary unit) after projecting to the eigenstates vs. $g$ for the 5-qubit (top panel (a)) and 6-qubit (bottom panel (b)) transverse-field Ising model. The procedure starts with two different initial states: (1) [(blue) dots that start on the lowest curve] the  ground state $|00000\rangle$ of $H_{\rm TFI}(g=0)$, and (2)  [(red) dots that start on the top curve] the highest-energy state $|11111\rangle$ of $H_{\rm TFI}(g=0)$.  }
		\label{fig:gIsingN56}
	\end{figure}
	
	Here, we return to the transverse-field Ising model~(\ref{eqn:Ising}) and perform the quantum annealing with our spectral projection as a subroutine. The ground state at $g=0$ is unique and is given by $|0^{\otimes N_q}\rangle$. But the ground states at $g=1$ are doubly degenerate, and they are $|+^{\otimes N_q}\rangle$ and $|-^{\otimes N_q}\rangle$. Similar to the previous section, we examine small system sizes with $N_q=5$ and $N_q=6$, shown in Fig.~\ref{fig:gIsingN56}. Given that there is small gap in both cases, the quantum annealing works.
	
	In the above simulations we have used the $e^{-i\hat{h}\Delta t}$ without decomposing it into Trotter terms. In order to simulate larger systems, we separate the Hamiltonian into two parts: $H_{\rm e}(g)$ and $H_{\rm o}(g)$ for even and odd bonds, where terms in $H_{\rm e}$ commute with one another and similarly for the terms in $H_{\rm o}$. Thus we can apply a Trotter-Suzuki decomposition to $e^{-i (H_{\rm o}+ H_{\rm e})\Delta t}\approx e^{-i H_{\rm o}\Delta t}e^{ -i H_{\rm e}\Delta t}$. This simulates the scenario that in the quantum circuit one can apply simultaneously the commuting terms of the controlled version of $e^{-i H_{\rm o}\Delta t}$ and subsequently those  of $e^{-i H_{\rm e}\Delta t}$. In our classical simulations, we use a 4-th order Trotter-Suzuki decomposition~\cite{Sornborger1999,DhandSanders} for $e^{-i (H_{\rm o}+H_{\rm e}) \Delta t}$: 
	\begin{eqnarray}
	&&e^{-i (H_{\rm o}+H_{\rm e}) \Delta t}\approx e^{-ia_1 H_{\rm o}\Delta t}e^{-i a_1 H_{\rm e}\Delta t}\\
	&&\qquad e^{i a_2 H_{\rm e}\Delta t}e^{i a_2 H_{\rm o}\Delta t}e^{-ia_3 H_{\rm o}\Delta t}e^{-ia_3 H_{\rm e}\Delta t}e^{-ia_3 H_{\rm e}\Delta t}\nonumber\\
	&&\qquad e^{-ia_3 H_{\rm o}\Delta t}e^{i a_2 H_{\rm o}\Delta t}e^{ia_2 H_{\rm e}\Delta t}e^{-ia_1 H_{\rm e}\Delta t}e^{-ia_1 H_{\rm o}\Delta t},\nonumber
	\end{eqnarray}
	where $a_1=(2+\sqrt{2})/4$, $a_2=-a_1$, and $a_3=(1+\sqrt{2})/2$.
	
\begin{figure}[t]
	(a)\\	\includegraphics[width=0.48\textwidth]{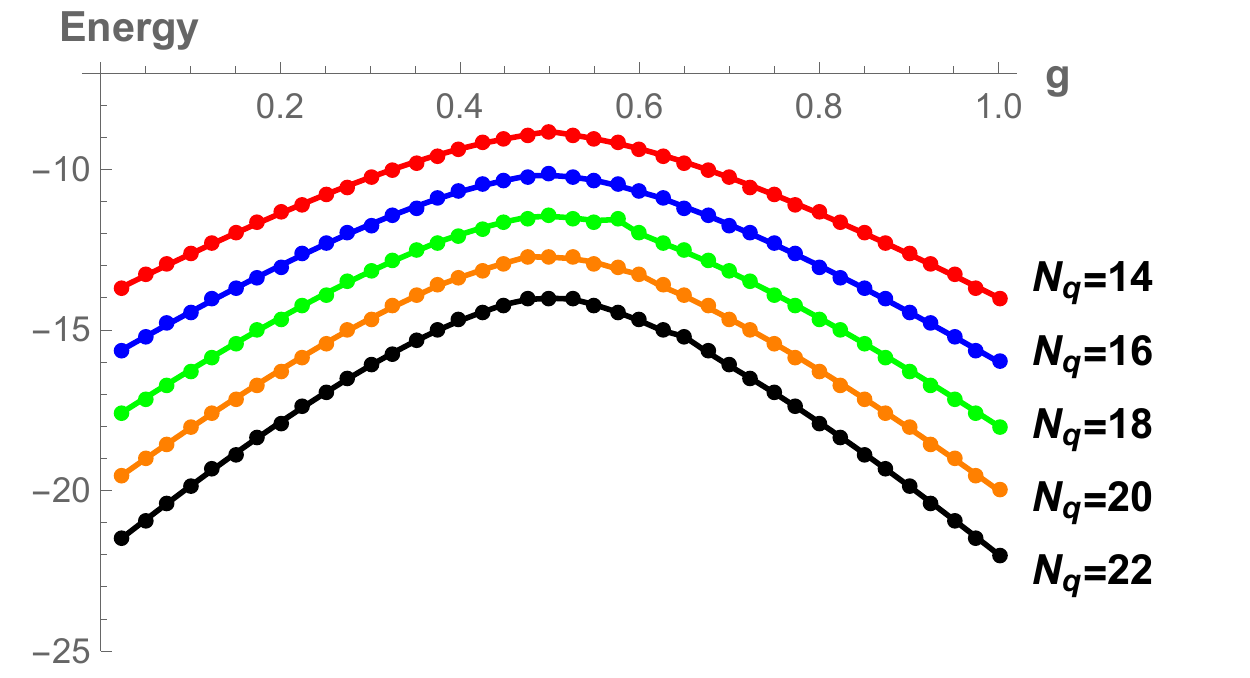}\\
	(b)\\	\includegraphics[width=0.48\textwidth]{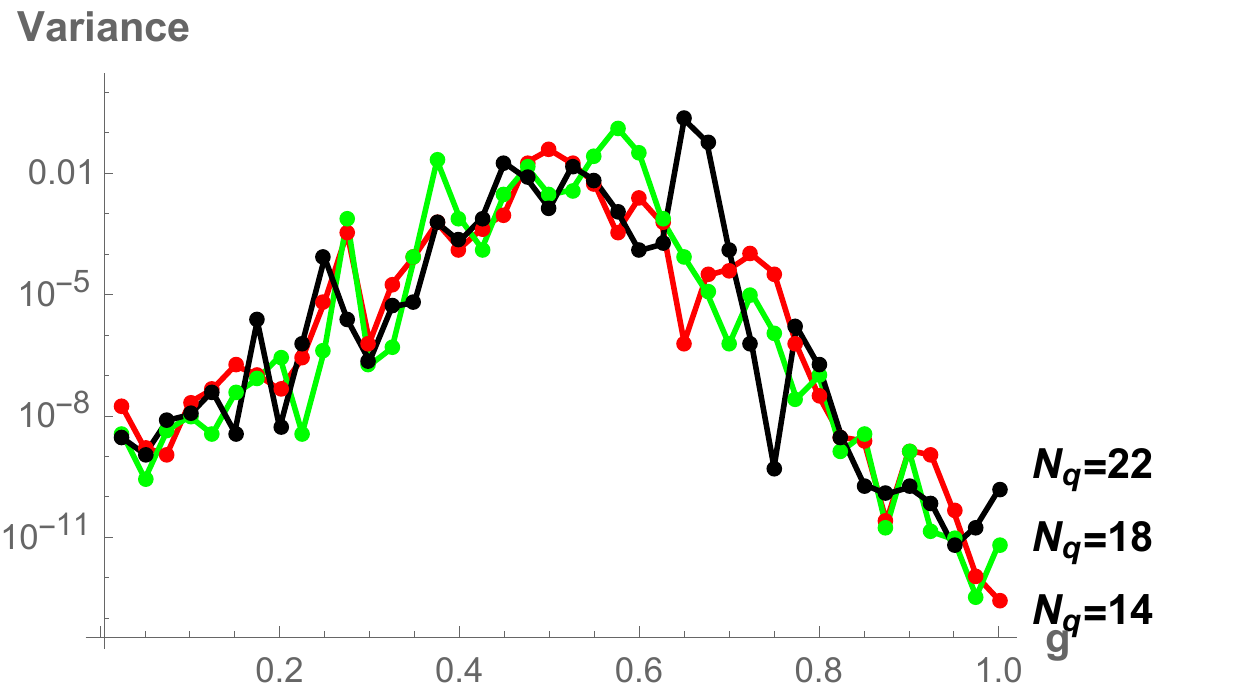}
		\caption{Application of our spectral projection algorithm in the quantum annealing as the subroutine that carries out the measurement to project to eigenstates ~\cite{Somma2008}. (a) The top figure shows the energy (in an arbitrary unit) after projecting to the eigenstates vs. $g$. Different colors represent different qubit numbers $N_q=14,16,18,20, \&22$. The procedure starts with the ground state $|00\cdots0\rangle$ of $H_{\rm TFI}(g=0)$. (b) The bottom figures shows the energy variance (in an arbitrary unit) at each step of projection (showing only for $N_q=14,18, \&22$ for illustration), which can be used as a figure of merit for the error in the energy. Generally, the variance is the largest around $g=0.5$, which is the critical point of the model in the thermodynamic limit. The curves are drawn to connect dots and to guide  the eye.
	  There are in total 210 steps in each projection run. The parameter $\Delta t$ is chosen from the list with number of repetitions shown in the parenthesis: $[0.01(\times10), 0.1(\times10), 0.03(\times50), 0.01(\times100), 0.003(\times40)]$ The ancillary state is chosen as $(\alpha=1/\sqrt{2},\beta=e^{i\phi}/\sqrt{2})$ with $\phi$ chosen randomly in $[0,2\pi)$.}
		\label{fig:EnVsG}
	\end{figure}

	As seen in Fig.~\ref{fig:EnVsG} with $N_q=114,16,18,20, \&22$, the annealing proceeds at initializing the state at the ground state of $H_{\rm TFI}(g=0)$, which is $|00\dots 0\rangle$. Then the spectral projection is applied successively at $g=j \Delta g$ for $j=1,2,\dots,40$ and $\Delta g=0.025$ (with the primitive being run 210 times in each projection procedure), ending at $g=1$ at the end of the annealing. The final energy after the anneal is seen to be close to the final ground-state energy, which is $-N_q$. The accuracy in this case can be increased by making the $\Delta g$ smaller and total number of iteration steps larger. The energy variance is generally the largest around $g=0.5$, and this is expected as, in the thermodynamic limit, there is a second-order quantum phase transition at $g_c=0.5$, and it is known that the gap closes as ${\cal O}(1/N)$ when $g$ approaches $g_c$ from below. As $g$ approaches 1, the ground state becomes doubly degenerate.

\subsection{Effect of decoherence in the annealing}
\begin{figure}[t]
	(a)\\	\includegraphics[width=0.48\textwidth]{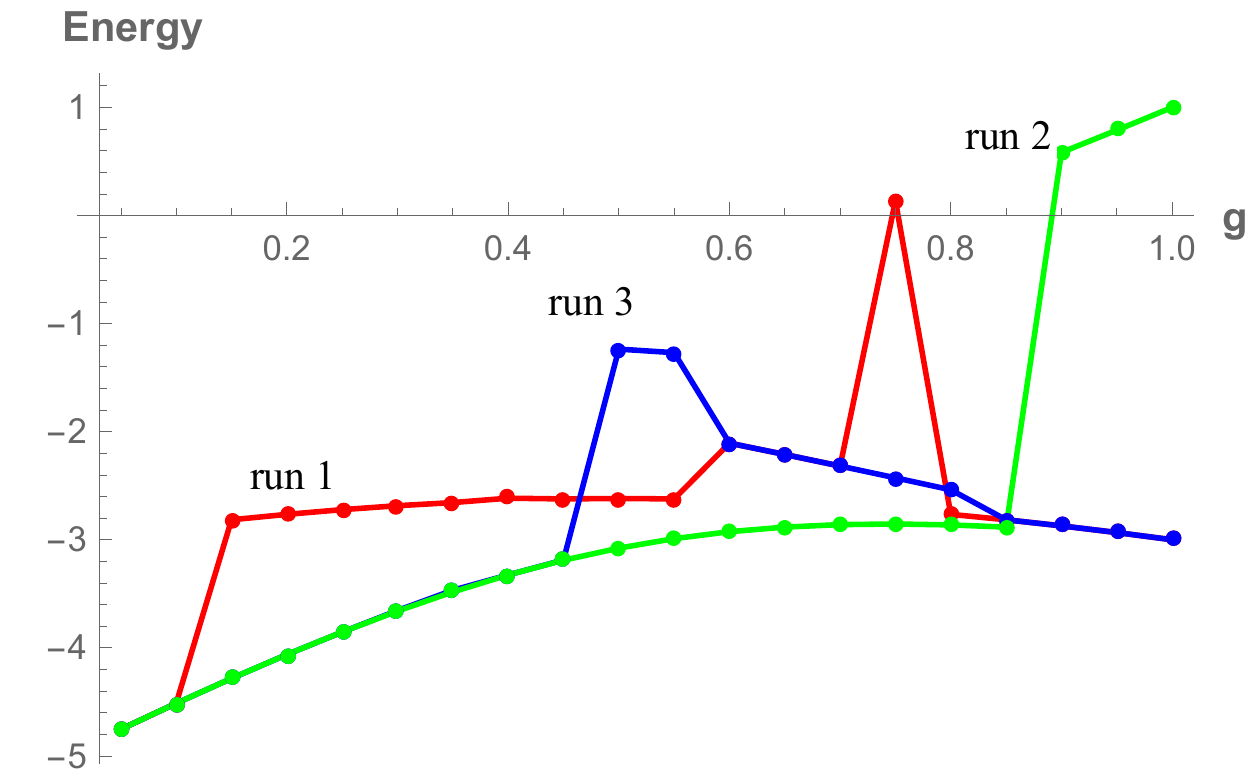}\\
		(b)\\\includegraphics[width=0.48\textwidth]{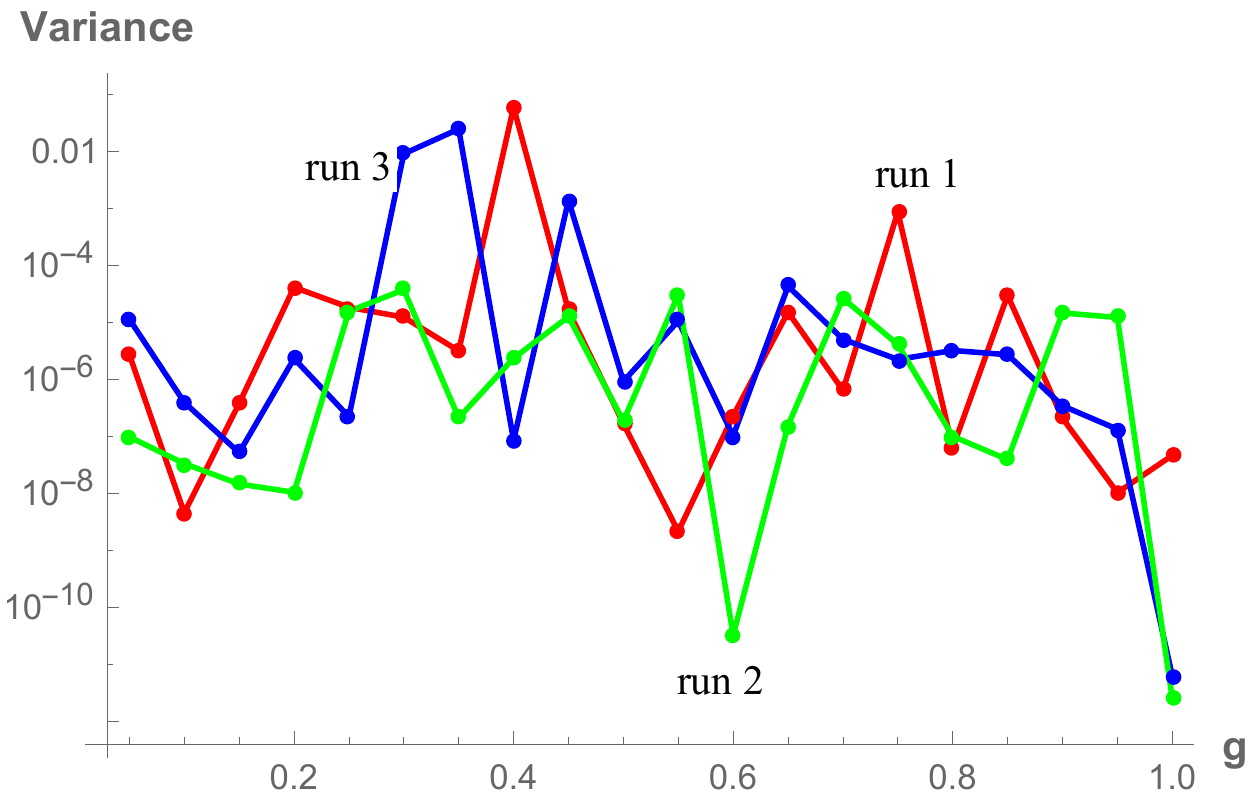}
		\caption{The effect of decoherence on the quantum annealing. We use the contrived scenario that the decoherence with $\epsilon=0.01$ occurs at every 31th step in our spectral projection subroutine, where the primitive is run for 180 steps. We use  the 5-qubit transverse-field Ising model ${H}_{\rm TFI}(g)$ and carry out 3 different runs. The energy (a) [top] and its variance (b) [bottom] are shown as 
		$g$ is varied. Both the energy and its variance are displayed in arbitary units. The initial state is the ground state $|00000\rangle$ of $H_{\rm TFI}(g=0)$. The final grounds are doubly degenerate and are $|+-+-+\rangle$ and $|-+-+-\rangle$.  }
		\label{fig:NoisyAnneal}
	\end{figure}

Here, we take into account of the decoherence effect in our spectral projection and discuss how it affects the quantum annealing. In general, our algorithm does not project against decoherence, as discussed in Sec.~\ref{sec:decoherence}, and hence the resulting quantum annealing will be worse than the noise-free case. The degree of inaccuracy depends on the error rate $\epsilon$. We use, as an illustration, the contrived scenario discussed above that the decoherence with $\epsilon=0.01$ occurs at every 31th step in our spectral projection subroutine, in which the primitive is run for 210 steps. We test this on the 5-qubit transverse-field Ising model ${H}_{\rm TFI}(g)$ and the results of three different runs for the quantum annealing are shown in Fig.~\ref{fig:NoisyAnneal}. As opposed to the noise-free case, there is some probability (depending on the noise rate and strength) that the final state may end up far from the final ground state. But there is also some probability that the final state is close to the final ground state. Developing noise-protecting spectral projection is thus a desirable goal that can yield a noise-protecting quantum annealing algorithm.

\begin{table}
\begin{tabular}{|l||l|l|l|}
\hline
{Methods}&  \begin{tabular}{l}Projection\cr capability\end{tabular} & \begin{tabular}{l}Phase\cr estimation\cr capability\end{tabular} & \begin{tabular}{l}Accuracy \cr limitation\end{tabular}\cr
\hline 
QPE & yes & yes & \begin{tabular}{l} no. of ancillas \& \cr power in $c-U^{2^{k}}$;\cr requires QFT\end{tabular} \cr
\hline
iQPE & yes & yes & \begin{tabular}{l} power in $c-U^{2^{k}}$;\cr requires no QFT\end{tabular}\cr
\hline
SPA  & yes & yes & \begin{tabular}{l} no. of iterations; \cr  requires no QFT\end{tabular}\cr
\hline
\end{tabular}
\caption{Comparison of the standard QPE~\cite{NielsenChuang}, the iterative QPE (iQPE)~\cite{Dobsicek2007} and our spectral projection algorithm (SPA). QFT stands for quantum Fourier transform.
Both the QPE and iQPE have fixed accuracy set by the choice of highest power in $U^{2^k}$ and during the procedure the highest power cannot be changed; the QPE is fixed by the total number of ancillas and the iQPE needs to fix the highest power in the beginning of the procedure. Both the QPE and iQPE require precise execution of $c-U^{2^k}$ for all $k< t_g$.
On the other hand, our SPA uses $c-e^{-i \hat{h}\Delta t}$ and the range of $\Delta t$ can be fixed, but the accuracy can still be improved by running more iterations. Our SPA does not require $\Delta t$ to be exact $2^k$, and in fact it can be somewhat arbitrary. The drawback of our SPA is that the number of required iterations for achieving a fixed accuracy can vary from run to run.
\label{tb}}
\end{table}
\section{Concluding remarks}
\label{sec:conclusion}
We have proposed a quantum algorithm for projecting to eigenstates of any Hermitian operator, provided one can access the associated control-unitary evolution and measurement of the controlling ancilla qubit. The procedure is iterative and the distribution of the projected eigenstates obeys the Born rule. It is robust against imprecision in timing. But it has only limited resilience against decoherence; the iterative procedure takes the system towards eigenstates, even after the influence of decoherence such as a depolarizing. It has no capability of error correction or prevention. We view our method as  a simpler algorithm to project the system into eigenstates of a Hermitian observable than the standard QPE and it can also be used to extract eigenvalues. We compare our spectral decomposition to the standard QPE and an iterative version in Table~\ref{tb}. Our algorithm can  be used as a subroutine in the quantum annealing 
procedure by measurement~\cite{Somma2008} to drive to the ground state of a final Hamiltonian. We have performed simulations that demonstrate the utility of our algorithm.   
We note that a previously proposed scheme of ground state cooling quantum computation also uses ancilla measurement for the cooling~\cite{GSCQC}. Our scheme uses ancilla measurement for the spectral projection and the way it is used in the QSA is similar to the quantum Zeno effect. It will be useful to develop a noise-protecting spectral projection. A  proof-of-principle demonstration of our spectral projection algorithm on currently available quantum computers will also be desirable.

Post-selection allows projection to the ground state, but the probability for obtaining the desired post-selected outcome is exceedingly small. The algorithm that we have attempted for the imaginary time evolution suffers some problems that make it not practical; see Appendix~\ref{app:imaginary-time}. The fact that we end up with a spectral projection that obeys the Born rule seems to indicate that we may need to go beyond the primitive used in this paper to achieve an imaginary-time evolution quantum algorithm, as done in Ref.~\cite{Motta2019}. But whether imaginary-time evolution can be achieved without using an effective Hamiltonian is an interesting question to consider.

In order to classically simulate our spectral projection algorithm, it will generally take exponential time in the number of qubits of the system, as one needs to compute $e^{-i\hat{h}\Delta t} |\psi\rangle$. Thus, it will be interesting if such a procedure can be carried out in a quantum computer for system sizes beyond the capability of classical simulations. This might be a useful playground for demonstrating quantum advantage.
\begin{acknowledgments}
This work was supported by National Science Foundation under grants No. PHY  1620252 and No. PHY 1915165. T.-C.W. acknowledges useful discussions with Fernando Brand\~ao, David Poulin and Barry Sanders. We also thank an anonymous referee for his/her suggestions that help improve the original manuscript.

\end{acknowledgments}
\appendix

\section{A failed attempt to construct an adaptive procedure for  imaginary-time evolution}
\label{app:imaginary-time}
	
	We have considered a primitive similar to the Hadamard test, except using an ancillary state of $\alpha|0\rangle+\beta|1\rangle$. The controlled unitary gate is of the form $c-e^{-i\hat{h}\Delta t}$. We consider the post-measurement state of the system up to the first order in $\Delta t$, 
	\be
|\psi'_m\rangle\approx \frac{1}{\sqrt{2}}	(\alpha+(-1)^m\beta)\left[1 - \frac{(-1)^m  i \Delta t}{\alpha/\beta+(-1)^m} \hat{h}\right] \ket\psi. \label{eq:measured}
	\ee
	
	Our motivation here is to achieve the nonunitary action $e^{-\hat{h}\Delta\tau}$ on $|\psi\rangle$, which, to first order, is $ [1-\hat{h}\Delta\tau] \ket\psi$. Let us choose to make it work for the $m=0$ outcome by requiring that
\be
\label{eqn:ratio}
	\frac{\alpha}{\beta}  = -1+ir, \ \ {\rm where}\,\, r \in \mathds{R},
	\ee
	then the nonunitary action is achieved, i.e., the effective action on the system is (ignoring normalization)
\be
\label{eqn:+outcome}
|\psi'_0\rangle\approx [1-\hat{h}\Delta t /r]|\psi\rangle,
\ee
obtaining an  effective time step $\Delta \tau=\Delta t/r$ in the imaginary-time evolution.
To satisfy Eq.~(\ref{eqn:ratio}), $\alpha$ and $\beta$ can be taken as  
	\begin{align}
	\label{eq:ab}
	\alpha(r)=\frac{-1+ir}{\sqrt{2+r^2}}, \
	\beta(r)=\frac{1}{\sqrt{2+r^2}}, \
	\end{align}
	and the probability for each outcome (without approximation)  is 
\begin{align}
\label{eqn:pm}
	p_m=\frac{1}{2}+\frac{(-1)^m}{2+r^2}\big(-{\rm Re}\langle \psi|U|\psi\rangle+ r \,{\rm Im}\langle \psi|U|\psi\rangle\big).
	\end{align}
	The Pauli X measurement on the ancilla can be realized by first applying the Hadamard gate $H$ before measuring in the standard Z basis; see Fig.~\ref{fig:oneIteration}.

However, for the outcome `1', the system will be collapsed to an undesired state, to the first order in $\Delta t$,
	\be
	|\psi'_1\rangle\approx\left[1 - i\frac{2}{r^2+4}\hat{h}\Delta t + \frac{r}{r^2+4}\hat{h}\Delta t\right] \ket\psi.
	\ee
The second term is not harmful, as by applying to the post-measurement state the `correcting'  unitary 
	\be
	U_{corr} = \exp{\left(i\frac{2}{r^2+4}\hat{h}\Delta t\right)},
	\ee
	the system becomes
	\be
	\label{eqn:-outcome}
	|\psi'_1\rangle\approx\left[1 + \frac{r}{r^2+4}\hat{h}\Delta t\right] \ket\psi,
	\ee
	to the first order in $\Delta t$. We note that this additional step is  not necessary as it only modifies the relative phases of different eigen-components, but not the amplitudes. 
	
	The second term inside the bracket of Eq.~(\ref{eqn:-outcome}) and Eq.~(\ref{eqn:+outcome}) represents the step size of a random walk in the exponent of an action $e^{-\hat{h}\Delta\tau_i}$ on a quantum state $|\psi\rangle$, where $\Delta\tau_0=\Delta t/r$ and $\Delta \tau_1=-\Delta t /(r+4/r)$ for the two respective measurement outcomes `0' and `1'; see Fig.~\ref{fig:oneIteration}b for illustration. The corresponding probabilities~(\ref{eqn:pm}) are approximately, 
	\begin{subequations}\label{eqn:p0}
	\begin{eqnarray}
	p_0(\psi) &\approx& \frac{r^2}{r^2+2} \left(\frac{1}{2} - \frac{h_{{\psi}}\Delta t}{r}\right), \\
	p_1(\psi)&\approx &\frac{r^2}{r^2+2}\left(\frac{1}{2} + \frac{h_{{\psi}}\Delta t}{r}\right) + \frac{2}{r^2+2}, 
	\end{eqnarray}
	\end{subequations}
where $h_{{\psi}} \equiv \bra{{\psi}}\hat{h}\ket{{\psi}}$ is the average energy for the state $\ket{{\psi}}$ of the system prior to this iteration. The dependence of $p$'s on the system state $|\psi\rangle$ prevents us from getting a closed-form expression for the outcomes of a long sequence of iterations.

	By post-selecting the `0' outcome in the primitive, and by repeating this one $n$ times we can achieve exponential decay to the ground state, via
\be
e^{-n{\Delta t}\,\hat{h}/{r}}|\psi\rangle.
\ee
 Imaginary time evolution is employed in many classical numerical methods, such as the iTEBD method for ground states~\cite{Vidal2004}.
However, for our quantum procedure the desired branch of having all `0' outcomes occurs with an exponentially small probability, so it is not very useful in practice. 

Instead of postselection, one may perform an additional operation if the undesired outcome `1' occurs.	We have attempted such idea but we did not succeed. What is described below is such a failed attempt.

\begin{figure}
		\includegraphics[width=0.5\textwidth]{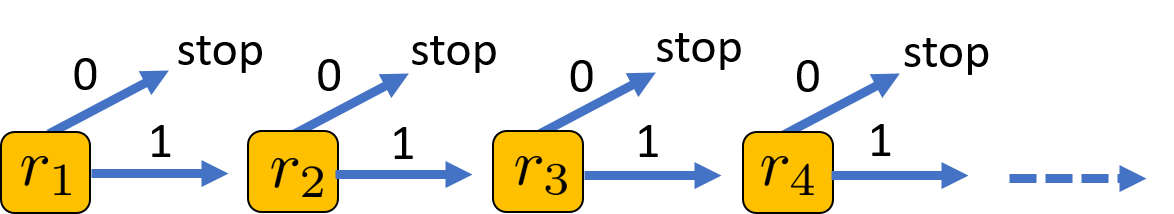}
		\caption{The diagram that illustrates the attempted algorithm for implementing one Trotter imaginary-time step.}
		\label{fig:oneStepIT}
	\end{figure}
Let us define one iteration to be the process from entangling the system with an ancilla to measuring the ancilla and possibly correcting with the unitary if needed. If the first step yields the `0' outcome, then one arrives at the desired imaginary-time evolution Eq.~(\ref{eqn:+outcome}). We ask what can one do if one obtains the `1' outcome and arrives  at a state in Eq.~(\ref{eqn:-outcome})? We can proceed with a second iteration by choosing a different parameter $r^\prime$.   The desired outcome `0' after this iteration would put the system in the state
	\be
	\label{eqn:1outcome}
	\left[1 + \frac{r}{r^2+4}\hat{h}\Delta t - \hat{h}\frac{\Delta t}{r^\prime}\right] \ket\psi.
	\ee
If we choose $r^\prime$ such that ${1}/{(r+4/r)}-{1}/{r^\prime}=-{1}/{r}$, i.e.,
\be
r^\prime=\frac{r}{2}\frac{r^2+4}{r^2+2},
\ee
then the outcome `0' leads to the desired imaginary-time evolution Eq.~(\ref{eqn:+outcome}).

However, if instead the measurement still gives the undesired outcome of `1', we need to correct it further by repeating the iteration until outcome `0' is obtained by choosing the parameter $r_{n+1}$ in the $(n+1)$-th round via  
\be
	r_{n+1} = \frac{r_n(r_n^2+4)}{2r_n^2+4},
	\ee
	and we terminate the iteration when `0' outcome is obtained. Then the desired one-step imaginary-time evolution will give
	\be
	[1-\hat{h}\Delta t /r_1]|\psi\rangle.
	\ee
	This procedure is summarized in Fig.~\ref{fig:oneStepIT}.
	
\begin{figure}[h]
		\includegraphics[width=0.48\textwidth]{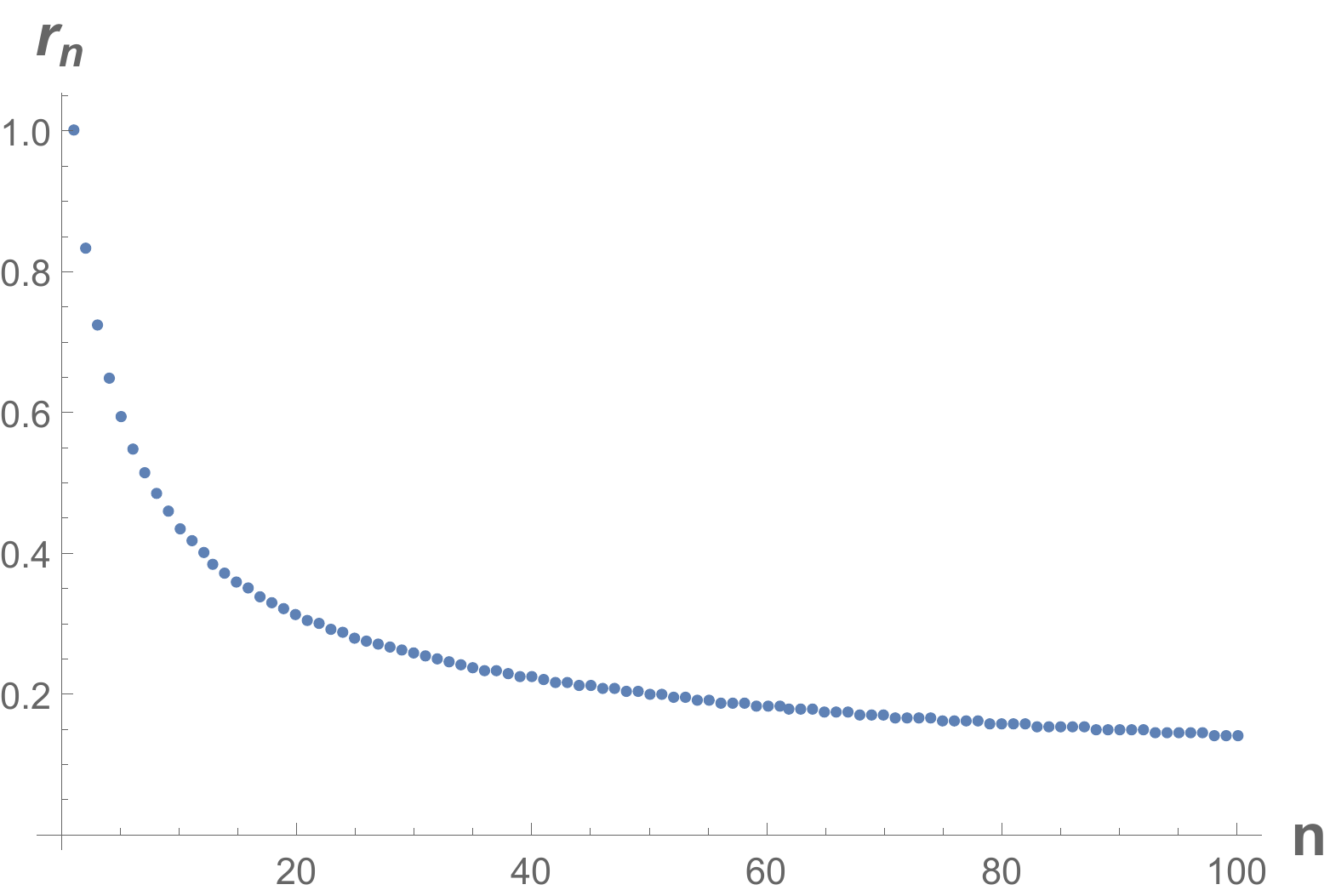}
		\caption{The first one hundred values of $r_n$, starting with $r_1=1$.}
		\label{fig:running_r}
	\end{figure}

	However, this procedure suffers from the occurrence of long sequences of `1' outcomes, as our simulations show. As a rough estimate by dropping the first-order contribution,  the probability of $n$ successive `1' outcomes is 
	\be
	p^{(n)} \equiv \prod_{j=1}^{n} p_1^{(j)} \approx \frac{r_{n+1}}{r_1},
	\ee
which does not decay exponentially. Figure~\ref{fig:running_r} shows the values of $r_n$ with $r_1=1$.  One can start with a larger $r_1$ so as to get a smaller ratio of $r_n/r_1$, but the scaling is still not exponentially small.

\section{Some derivations}
\label{app:derivations}
\subsection{Energy change}
Let us list the post-measurement state,
	\begin{align}
|\psi'_m\rangle
= \frac{1}{\sqrt{2}}[\alpha+(-1)^m\beta e^{-i\hat{t}\Delta t}] \ket\psi, 
	\end{align}
and the probability that it occurs,
	\begin{eqnarray}
	p_m&=&\frac{1}{2}\big[1+2(-1)^m {\rm Re}(\alpha^*\beta e^{-i\xi} \langle \psi|e^{-i\hat{h}\Delta t}|\psi\rangle)\big]\\
	&=&\frac{1}{2}\big[1+2(-1)^m {\cal R}_1\big],
	\end{eqnarray}
where it is convenient to define ${\cal R}_1$ and a related ${\cal R}_h$:
\begin{eqnarray}
{\cal R}_1&\equiv& {\rm Re}\big(\alpha^*\beta \langle \psi|e^{-i\hat{h}\Delta t}|\psi\rangle\big), \\
{\cal R}_h&\equiv& {\rm Re}\big(\alpha^*\beta \langle \psi|e^{-i\hat{h} \Delta t}\hat{h}|\psi\rangle\big). 
\end{eqnarray}
Thus the change in energy is
\begin{eqnarray}
&&\Delta E_{(m)}=\frac{1}{p_m}\langle\psi'_m |\hat{h}|\psi'_m\rangle -\langle\psi| \hat{h}|\psi\rangle\\
&&=\frac{\langle\hat{h}\rangle + 2(-1)^m {\rm Re}
\big(\alpha^*\beta \langle \psi|e^{-i\hat{h} \Delta t}\hat{h}|\psi\rangle\big) 
 }{1+2(-1)^m {\cal R}_1}-\langle \hat{h}\rangle\\
&&=
\frac{2(-1)^m\big({\cal R}_h-\langle h\rangle {\cal R}_1\big)}{1+2(-1)^m {\cal R}_1}.
\end{eqnarray}
Then, expanding  the above expression in series of $\Delta t$ is straightforward. 
\subsection{Change in average energy variance}
In the main text, we have the expression for the average energy variance
\begin{eqnarray}
\overline{\delta V_E}&=& \langle\psi|\hat{h}|\psi\rangle^2-\sum_{m=0,1}\frac{\langle{\psi}'_m| \hat{h}|{\psi}'_m\rangle^2}{p_m}\\
&=& \langle\hat{h}\rangle^2-\sum_{m=0,1}
\frac{\big(\langle\hat{h}\rangle+2(-1)^m {\cal R}_h\big)^2}
{2+4(-1)^m {\cal R}_1}
\end{eqnarray}
By expanding the square and explicitly summing over $m$, we obtain
\begin{equation}
\overline{\delta V_E}=\frac{-4}{1-4 {\cal R}_1^2}\big({\cal R}_1\langle h\rangle - {\cal R}_h\big)^2.
\end{equation}
Then, expanding  the above expression in series of $\Delta t$ is straightforward. 
\subsection{Average $\hat{Q}$ action}
By expanding the post-measurement state $|\psi'_m\rangle$ to the second order in $\Delta t$, we have
	\begin{eqnarray}
|\psi'_m\rangle\approx \frac{\alpha+(-1)^m\beta}{\sqrt{2}}\left[1+\frac{-i\hat{h}\Delta t -\frac{1}{2}(\hat{h}\Delta t )^2}{1+(-1)^m\alpha/\beta}\right]|\psi\rangle.\nonumber
	\end{eqnarray}
The goal is to the above equation to the exponentiated form 
	$|\psi'_m\rangle\sim e^{\hat{P}_m} |\psi\rangle$, correct to the second order. Naturally, $\hat{P_m}$ will contain the second term in the square bracket. But we also need to take into account of other contribution to the second order. So we can set
\begin{equation}
\hat{P}_m=\frac{-i\hat{h}\Delta t -\frac{1}{2}(\hat{h}\Delta t )^2}{1+(-1)^m\alpha/\beta}+ \hat{X}(\Delta t)^2.
\end{equation}
Expanding $e^{\hat{P}_m}$, we have to the second order,
	\begin{eqnarray*}
1+\frac{-i\hat{h}\Delta t -\frac{1}{2}(\hat{h}\Delta t )^2}{1+(-1)^m\alpha/\beta}
+\hat{X}(\Delta t)^2
	-\frac{1}{2}\frac{(\hat{h}\Delta t )^2}{[1+(-1)^m\alpha/\beta]^2}, \nonumber
\end{eqnarray*}
which should equals
\begin{align}
1+\frac{-i\hat{h}\Delta t -\frac{1}{2}(\hat{h}\Delta t )^2}{1+(-1)^m\alpha/\beta}. \nonumber
\end{align}
Therefore, we obtain
	\begin{equation}
	\hat{P}_m= \frac{-i\hat{h}\Delta t -\frac{1}{2}(\hat{h}\Delta t )^2}{1+(-1)^m\alpha/\beta}+\frac{1}{2}\frac{(\hat{h}\Delta t )^2}{[1+(-1)^m\alpha/\beta]^2}.
	\end{equation}
From this, it is straightforward to obtain $\hat{Q}_m=(\hat{P}_m+\hat{P}_m^\dagger)/2$ and perform the sum $\sum_{m=0,1}p_m\hat{Q}_m$. In the end, we arrive at
\begin{align}\sum_{m=0,1}p_m \hat{Q}_m=-\frac{{\rm Im}(\alpha^*\beta)^2 \Delta t^2}{1-4 {\rm Re}(\alpha^*\beta)^2}[(\hat{h}-\langle \hat{h}\rangle)^2 - \langle \hat{h}\rangle^2].
\end{align}
Then, expanding  the above expression in series of $\Delta t$ is straightforward. 

\end{document}